\documentclass[%
preprint,
superscriptaddress,showpacs,
 amsmath,amssymb,
 aps,
]{revtex4-2}

\usepackage{xcolor}

\usepackage{graphicx}
\usepackage{dcolumn}
\usepackage{bm}
\usepackage{fancyhdr}
\usepackage[english]{babel}
\usepackage[utf8]{inputenc}
\usepackage[colorlinks=true,
            linkcolor=blue,
            urlcolor=blue,
            citecolor=blue]{hyperref}

\begin{document}

\preprint{}

\title{Compartmentalizing the cuprate strange metal}

\author{M. Berben}
\thanks{These two authors contributed equally}
\affiliation{High Field Magnet Laboratory (HFML-EMFL) and Institute for Molecules and Materials, Radboud University, Nijmegen, Netherlands}
\thanks{These two authors contributed equally}

\author{J. Ayres}
\thanks{These two authors contributed equally}
\affiliation{H. H. Wills Physics Laboratory, University of Bristol, Bristol, United Kingdom}

\author{C. Duffy}
\affiliation{High Field Magnet Laboratory (HFML-EMFL) and Institute for Molecules and Materials, Radboud University, Nijmegen, Netherlands}

\author{R. D. H. Hinlopen}
\affiliation{H. H. Wills Physics Laboratory, University of Bristol, Bristol, United Kingdom}

\author{Y.-T. Hsu}
\affiliation{High Field Magnet Laboratory (HFML-EMFL) and Institute for Molecules and Materials, Radboud University, Nijmegen, Netherlands}

\author{M. Leroux}
\affiliation{LNCMI-EMFL, CNRS UPR3228, Univ. Grenoble Alpes, Univ. Toulouse, INSA-T, Grenoble and Toulouse, France}

\author{I. Gilmutdinov}
\affiliation{LNCMI-EMFL, CNRS UPR3228, Univ. Grenoble Alpes, Univ. Toulouse, INSA-T, Grenoble and Toulouse, France}

\author{M. Massoudzadegan}
\affiliation{LNCMI-EMFL, CNRS UPR3228, Univ. Grenoble Alpes, Univ. Toulouse, INSA-T, Grenoble and Toulouse, France}

\author{D. Vignolles}
\affiliation{LNCMI-EMFL, CNRS UPR3228, Univ. Grenoble Alpes, Univ. Toulouse, INSA-T, Grenoble and Toulouse, France}

\author{Y. Huang}
\affiliation{Van der Waals-Zeeman Institute, University of Amsterdam, Amsterdam, Netherlands}

\author{T. Kondo}
\affiliation{Institute for Solid State Physics, University of Tokyo, Kashiwa, Japan}

\author{T. Takeuchi}
\affiliation{Toyota Technological Institute, Nagoya 468-8511, Japan}

\author{J. R. Cooper}
\affiliation{Department of Physics, University of Cambridge, Cambridge, United Kingdom}

\author{S. Friedemann}
\affiliation{H. H. Wills Physics Laboratory, University of Bristol, Bristol, United Kingdom}

\author{A. Carrington}
\affiliation{H. H. Wills Physics Laboratory, University of Bristol, Bristol, United Kingdom}

\author{C. Proust}
\affiliation{LNCMI-EMFL, CNRS UPR3228, Univ. Grenoble Alpes, Univ. Toulouse, INSA-T, Grenoble and Toulouse, France}

\author{Nigel E. Hussey}
\affiliation{High Field Magnet Laboratory (HFML-EMFL) and Institute for Molecules and Materials, Radboud University, Nijmegen, Netherlands}
\affiliation{H. H. Wills Physics Laboratory, University of Bristol, Bristol, United Kingdom}

\date{\today}

\begin{abstract}
\textbf{Understanding the \lq anomalous normal' state from which pairs form and condense is key to unlocking the mystery of cuprate high-$T_c$ superconductivity. While many of its signature transport properties -- linear-in-temperature ($T$) resistivity, linear-in-field ($H$) magnetoresistance (MR) at high field strengths, quadratic-in-temperature inverse Hall angle and modified Kohler scaling -- have been identified, they are often considered either at a singular doping level or across a narrow doping range. As a result, their relation to each other and to the pseudogap, strange metal and non-superconducting regimes that define the cuprate phase diagram has yet to be elucidated. Here, we report an extensive series of high-field in-plane MR measurements on several cuprate families spanning all three regimes that reveal a systematic evolution of the MR, with each regime possessing its own distinct scaling behavior. In the pseudogap regime, the MR exhibits pure $H/T^2$ scaling and $H$-linearity at the highest field strengths. While the $H$-linearity persists inside the strange metal regime, the scaling changes abruptly to $H/T$. The magnitude of the $H$-linear slope is found to be correlated with both the $T$-linear resistivity coefficient and $T_c$, strengthening the characterization of the strange metal regime as a quantum critical phase. Finally, within the non-superconducting, Fermi-liquid regime, we observe a recovery of conventional Kohler scaling, highlighting the anomalous nature of the scaling in cuprates that superconduct. This comprehensive study -- over an unprecedented temperature, field and doping range -- enables us to compartmentalize the phase diagram into distinct regimes, each with its own scaling behavior and identifies power-law scaling of the normal state MR as a defining feature of superconducting hole-doped cuprates. Phenomenology that obeys such simple scaling laws suggests an equally simple organizing principle at play. The incompatibility of such power-law scaling with any known variant of Boltzmann transport theory, however, motivates the quest for an altogether new theoretical framework, one in which the MR is entirely decoupled from all elastic scattering processes.}
\end{abstract}


\maketitle

\date{\today}

\maketitle

\section{Introduction}
The anomalous normal-state transport properties of high-$T_c$ cuprates \cite{gurvitch_prl_1987, Martin_PRB_1990, Chien_PRL_1991, Ando_PRL_1995, harris_prl_1995, fournier98, marel_2003, cooper_science_2009, Jin_Nature_2011, badoux_2016, giraldogallo_science_2018, Ayres_2021, yuan_2021} continue to challenge our understanding of how metallicity and superconductivity emerge in a two-dimensional doped Mott insulator \cite{Varma_PRL_1989, Anderson_PRL_1991, Emery_1995, zaanen_nature_2004, wu_2021}. Despite decades of concerted effort, a consensus has yet to be reached on how the various phases/regimes that form the cuprate phase diagram evolve or indeed relate to one another. At sufficiently high (electron or hole) doping, it is now well established that the system recovers a conventional, albeit highly correlated Fermi-liquid (FL) ground state characterized by a pure $T^2$ zero-field resistivity $\rho(0,T)$ at low temperatures \cite{Manako_PRB_1992, nakamae_2003, Jin_Nature_2011}. Identifying the doping level $p$ at which the FL ground state is restored, however, has proved highly controversial. In the case of hole-doped cuprates, this controversy centers on two prominent, though distinct scenarios. 

The first is a conventional quantum critical scenario in which  the FL ground state is restored beyond a specific hole doping $p^{\ast} \sim 0.2$ where the normal state pseudogap (a partial gap in the electronic density of states) closes \cite{keimer_quantum_2015, varma20}. Resistivity curvature mapping -- plots of the second derivative d$^2\rho$/d$T^2$ as a function of $T$ and $p$ -- was used early on to trace out the crossover line $T^*(p)$ (for $p < p^*$) below which manifestations of pseudogap formation first appear in the dc transport \cite{Ando_2004_curvature}. $T^*(p)$ also marks the temperature above which $\rho(0,T)$ adopts its signature $T$-linearity extending up to the highest accessible temperatures \cite{gurvitch87, martin90, bucher93}. As $p$ increases, $T^*(p)$ drops monotonically before vanishing abruptly in the vicinity of $p^{\ast}$ \cite{Hussey_PhilTrans_2011}. Beyond $p^{\ast}$, a second temperature scale $T_{coh}$ emerges, above which $\rho(0,T)$ is again proportional to $T$ \cite{Hussey_PhilTrans_2011, Kaminski_PRL_2003}. Together, these two scales define a fan-shaped region of the phase diagram, centred on $p^{\ast}$ and hosting a robust $T$-linear resistivity. Its resemblance to a quantum critical fan, coupled with putative evidence for multiple broken symmetries below $T^*$ (\cite{zhao17} and references therein), has motivated interpretations of $T^*$ as a second-order phase transition line terminating at $p^{\ast}$ \cite{keimer_quantum_2015, michon19, varma20}.

A key characteristic of the $T$-linear resistivity near $p^*$ is the magnitude of its slope ($\sim$ 1-2 $\mu\Omega$cm/K) that can be parameterized in terms of an effective scattering rate $\hbar/\tau \sim \zeta k_BT$ with $\hbar$ Planck's constant, $k_B$ Boltzmann's constant and $\zeta$ a dimensionless number 1 $< \zeta < \pi$ \cite{Bruin_science_2013, legros_natphys_2019, harada_2021}. This rate is widely (though not universally \cite{sadovskii, huang_2019}) attributed to Planckian dissipation \cite{zaanen_nature_2004}, a notional bound on the equilibration rate for charge carriers in correlated metals \cite{hartnoll_mackenzie}. While there is as yet no microscopic theory for Planckian dissipation, it is generally attributed to scattering off quantum critical fluctuations associated with the end of the pseudogap regime.

One of the major challenges to any scenario based on conventional quantum criticality is the persistence of a $T$-linear component in $\rho(T)$ (= $a_1T$) far beyond $p^\ast$ and down to the lowest temperatures studied (currently $\sim$ 0.02 K for electron-doped \cite{Jin_Nature_2011} and $\sim$  0.1 K for hole-doped cuprates \cite{Mackenzie_prb_1996, Proust_prl_2002}). Such \lq strange metallic' behavior \cite{phillips22} implies that a FL ground state is {\it not} restored at $p^{\ast}$. Rather, a second scenario has been proposed; that of a quantum critical {\it phase} extending from $p^\ast$ to $p_{sc} \sim 0.3$ -- the doping level corresponding to the edge of the superconducting (SC) dome \cite{cooper_science_2009, hussey_tale_2018, legros_natphys_2019}.

Deducing which scenario is valid is important as it may ultimately determine whether the transition into the SC state beyond optimal doping involves coherent quasiparticles $\grave{a}$ $la$ BCS (possibly with an effective mass enhanced by critical fluctuations) or whether it requires a more exotic non-BCS framework \cite{pelc_2019, culoduffy, inkof22}.

On a fundamental level, such limiting low-$T$ $T$-linear resistivity can arise from scattering off bosonic fluctuations with a characteristic frequency $\Omega$ that is lower than the temperature scale \cite{seibold_2021}. While charge and spin fluctuations have been extensively probed in overdoped cuprates \cite{Wakimoto_2007, Li_2018, Arpaia_2019, Miao_2021, Li_2021, Tam21}, there is as yet no evidence that $\Omega \rightarrow$ 0 or that the correlation length of the associated order parameter diverges, neither at $p^{\ast}$ nor at any doping level up to $p_{sc}$. This lack of an identifiable quasi-static bosonic mode has compelled theorists to look beyond conventional many-body approaches \cite{Blake15, patel18, zaanen_scipost_2019} and experimentalists to look beyond $\rho(0,T)$ \cite{giraldogallo_science_2018, Husain_2019, Putzke_NatPhys_2021, Ayres_2021, grissonnanche_2021} for additional clues as to the nature of the strange metal.

In principle, magnetoresistance (MR) provides complementary information about the nature of the scattering processes underlying $\rho(0,T)$. Moreover, even in systems that appear highly complex, important insights can often be gained through the observation of simple scaling laws. The low-field MR of many conventional metals, for example, obeys Kohler scaling:
\begin{equation}
    \Delta \rho(H,T)/\rho(0,T) = f(H/\rho(0,T))
    \label{eq:Kohler}
    \end{equation}
where typically, $f(H/\rho(0,T)) \propto (H/\rho(0,T))^2$ (the low-field MR being quadratic in field strength). Adherence to Kohler scaling implies that the $T$-dependence of the MR is (almost) entirely governed by changes in $\hbar/\tau$ with temperature or purity. Implicit in this rule is the fact that the scattering processes that determine $\rho(0,T)$ are the same as those that set the size of the MR itself.  

In cuprates, even those comprising a single, cylindrical Fermi sheet, conventional Kohler scaling breaks down. Near optimal doping, for example, the low-field MR is claimed to follow a \lq modified' Kohler scaling \cite{harris_prl_1995}:
\begin{equation}
    \Delta \rho(H)/\rho(0) =  s(\mathrm{tan^2}\theta_\mathrm{H})
    \label{eq:MKR}
    \end{equation}
where the Hall angle $\theta_\mathrm{H} \equiv \mathrm{tan}^{-1}(\rho_{xy}/\rho_{xx})$ and $s$ is a prefactor. Since  $\mathrm{cot}\theta_\mathrm{H} \sim A + BT^2$ \cite{Chien_PRL_1991} and $\rho(T) \propto T$, modified Kohler scaling has often been interpreted as a manifestation of the separation of transport and Hall lifetimes \cite{Anderson_PRL_1991},  the $T^{-1}$/$T^{-2}$ lifetimes being associated with relaxation processes normal/tangential to the Fermi surface in response to the application of an electric/magnetic field, respectively. We note that similar behavior has now been reported in other correlated electron systems \cite{Nakajima07, Kasahara10, Huang_PRR_2020}.

Recent MR studies -- carried out to much higher field strengths -- have uncovered an altogether different phenomenology however. The quadratic MR seen at low fields is found to evolve into a $H$-linear MR (LMR) with a $T$-dependent crossover field. Such behavior is best captured by the empirical expression:
\begin{equation}
    \rho(H, T) = \mathcal{F}(T)+\sqrt{(\alpha T)^2 + (\gamma \mu_0 H)^2}~,
    \label{eq:QuadT_2}
\end{equation}
where $\mathcal{F}(T)$ is a field-independent contribution to the resistivity and $\alpha$ and $\gamma$ are fixed coefficients. This form of MR, first identified \cite{Hayes_NatPhys_2016} in BaFe$_2$(As$_{1-x}$P$_x$)$_2$ near its magnetic QCP (with $\mathcal{F}(T) \simeq \rho_0$, the residual resistivity), has since been observed in the iron chalcogenides \cite{licciardello_coexistence_2019} as well as in highly-doped cuprates \cite{Ayres_2021}. The coupling of $H$ and $T$ in quadrature leads to $H/T$ scaling, in which a series of MR sweeps plotted as $\Delta \rho/T$ = ($\rho(H,T) - \mathcal{F}(T))/T$ vs.~$H/T$ collapse onto a single curve. Eq.~(\eqref{eq:QuadT_2}) also implies that at low-$T$, the LMR has a slope that is $T$-independent and set by $\gamma$ and in zero-field, $\rho(0,T)$ possesses a strictly $T$-linear component $\alpha T$.

There are several elements of Eq.~(\eqref{eq:QuadT_2}) that stand out. The first is that the residual (elastic scattering) contribution is contained only in $\mathcal{F}(T))$ and not in the MR. This aspect of the scaling is wholly incompatible with any form of Kohler scaling, conventional or otherwise, in which the $T$-dependence of the MR is set by the \textit{total} scattering rate. The second is the implicit association of this form of MR scaling with $T$-linear zero-field resistivity. In the pnictides and chalcogenides, this association has tied the quadrature MR to their respective QCPs. In highly-doped cuprates, however, it establishes a link between the $H/T$ scaling seen far beyond $p^{\ast}$ \cite{Ayres_2021} and the persistent $T$-linear component in $\rho(0,T)$ \cite{cooper_science_2009, hussey_2013, legros_natphys_2019}. The final notable feature of Eq.~(\eqref{eq:QuadT_2}) is that $\mathcal{F}(T))$ can also include a $T$-dependent term. This feature, most evident in the overdoped cuprates \cite{Ayres_2021}, implies that $\rho(H,T)$ can be decomposed into two distinct components, only one of which has a non-negligible MR. 

The origin of this dual character of the dc transport is not known at present. Moreover, while that initial study \cite{Ayres_2021} reaffirmed the notion that a FL ground state is \textit{not} restored at $p^{\ast}$, questions remained about the universality of the phenomenology. The purpose of this article is to examine the evolution of the in-plane MR across the entire phase diagram of hole-doped cuprates and to establish the experimental framework out of which further theoretical understanding can emerge. In so doing, we seek to address the following open questions:

\textbf{A}: How does the magnitude of the high-field $H$-linear slope $\gamma$ evolve across the SC dome? Specifically, how does it relate to $T_c$, $p$ or $a_1$ -- the $T$-linear component of the low-$T$ zero-field resistivity? While the correlation of the latter with $T_c$ is well established \cite{cooper_science_2009, yuan_2021}, the inclusion of $\mathcal{F}(T)$ in Eq.~(\eqref{eq:QuadT_2}) cautions that $\alpha$ and $a_1$ are not necessarily equivalent. The evolution of $\gamma(p)$ should indicate whether or not their origins are intimately linked.

\textbf{B}: How does the MR scaling evolve below $p^{\ast}$? If Kohler scaling is found to emerge within the pseudogap regime, as previously suggested \cite{chan_prl_2014}, one could map how this FL component emerges out of the strange metal regime. If, on the other hand, the power-law scaling is preserved, this would highlight a distinct non-FL component to the remnant carriers below $p^{\ast}$.

\textbf{C}: What happens beyond the SC dome? More specifically, does the loss of superconductivity on the overdoped side coincide with the disappearance of LMR at high fields and/or the recovery of a more conventional Kohler scaling? This question links directly to the controversy highlighted above surrounding the exact location (in $p$) of the emergence of a FL ground state in overdoped cuprates. 

\textbf{D}: How can $H/T$ scaling be reconciled with the modified Kohler scaling seen at low fields? While the distinct $T$-dependencies of the zero-field and low-field or Hall resistivities have been an essential ingredient of many phenomenological models of the cuprate normal state \cite{Anderson_PRL_1991, Carrington92, Stojkovic97, Ioffe98, Hussey03, Barisic13, Rice_PRB_2017}, all such models were proposed \textit{before} the discovery of power-law MR scaling. 

In order to address these outstanding questions, we have carried out a detailed study of the high-field MR of three different families of single-layered cuprates: LSCO, (Pb/La)-doped Bi$_2$Sr$_2$CuO$_{6+\delta}$ (Bi2201) -- covering a far wider doping range than previous studies \cite{giraldogallo_science_2018, Ayres_2021} -- and Tl$_2$Ba$_2$CuO$_{6+\delta}$ (Tl2201) under hydrostatic pressure, combined with analysis of existing data on Bi2201, LSCO and HgBa$_2$CuO$_{4+\delta}$ (Hg1201). The breadth of our study, over unprecedented field, temperature and doping/pressure ranges, has enabled us to provide definitive answers to all of the above questions, as we outline below. At the same time, the abrupt changes observed in the form of MR scaling across the phase diagram allow us to compartmentalize the magnetotransport of hole-doped cuprates into three distinct doping regimes: the pseudogap, the strange metal and the non-superconducting FL regime. Such compartmentalization supports existing models characterizing the pseudogap as a doping-dependent energy scale that collapses at $p^{\ast}$ \cite{Tallon_PRB_2020, Nachtigal_20}, rather than as a temperature scale associated with a transition into some form of hidden order. The most striking observation, however, is the crossover from conventional Kohler scaling in the FL regime to pure power-law scaling for all superconducting samples, the latter being wholly incompatible with Boltzmann transport theory. Overall, these findings reveal that the normal state of high-$T_c$ superconductors has its own unique magnetic field response, one that is seemingly decoupled from all elastic scattering processes.

\begin{figure}[!ht]
              \centering        
              \includegraphics{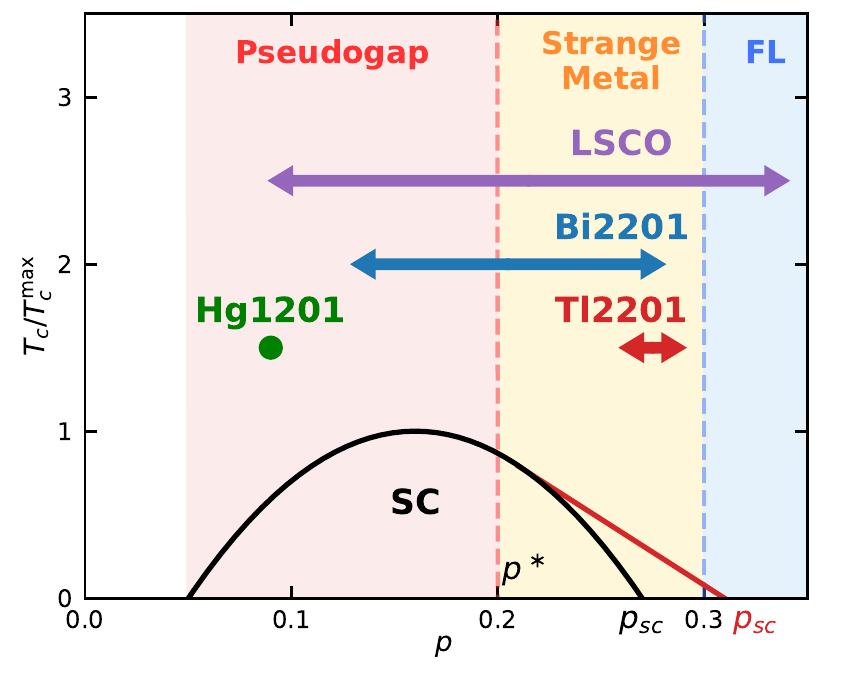}
              \caption{Simplified phase diagram of hole-doped cuprates showing the doping ranges (as double-headed arrows) of different families studied in this work and elsewhere \cite{Kimura96, harris_prl_1995, giraldogallo_science_2018, cooper_science_2009, chan_prl_2014}. Dashed lines indicate the boundaries (in $p$ at $T$ = 0) between the pseudogap, strange metal and FL regimes, as inferred from the magnetotransport. The solid line represents the superconducting (SC) dome for LSCO and Bi2201 (black line) and Tl2201 (red line), respectively, normalized to $T_c^{\rm max}$ -- the maximum value of $T_c$ -- in each family.}
              \label{fig:phase-diagram}
\end{figure}
 

\section{Experimental Survey}

\begin{figure*}[!ht]
              \includegraphics{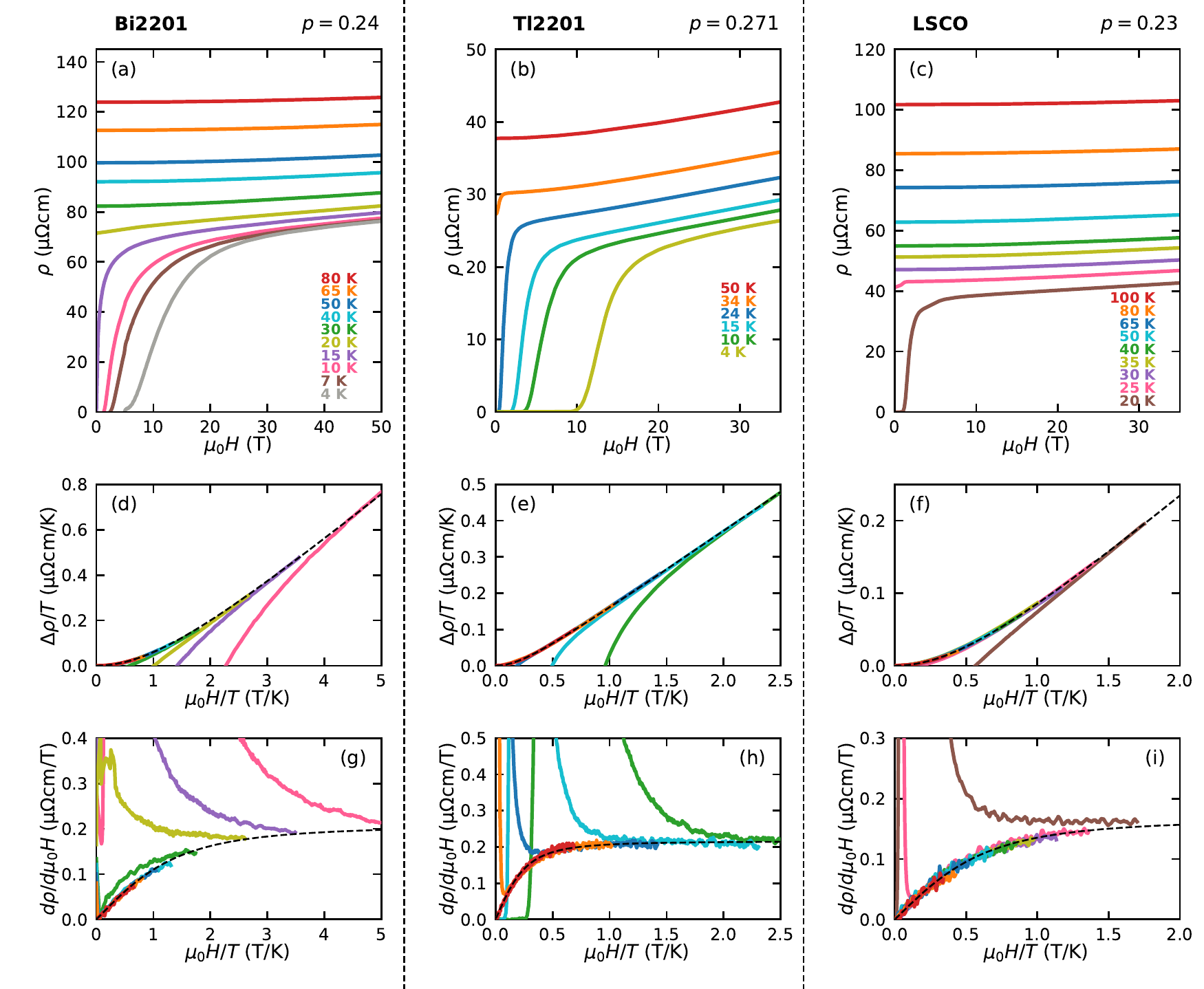}
              \caption{(a)-(c) Representative $\rho(H, T)$ curves for cuprates within the strange metal regime: (a) Bi2201 $p$ = 0.24, (b) Tl2201 $p$ = 0.275 and (c) LSCO $p$ = 0.23. Note the crossover to $H$-linear MR at the highest fields with a slope that is temperature independent. (d)-(f) Demonstration of adherence to $H/T$ (quadrature) MR scaling after subtraction of the zero-field resistivity, i.e. $\Delta \rho = \rho(H,T) - \rho(0,T)$, for the same three samples. For MR curves below $T_c$, $\rho(0,T)$ was obtained from universal fits to Eq.~(\ref{eq:QuadT_2}). (g)-(i) Corresponding scaling of the derivatives d$\rho$/d$\mu_0 H$. The data in panel (g) are reproduced from Ref.~\cite{Ayres_2021}. Dashed lines in all panels (d)-(i) are fits to the derivative of the quadrature form given by Eq.~(\ref{eq:QuadT_2}).}
              \label{fig:data-summary}
\end{figure*}

In order to preempt our subsequent discussion, we introduce in Figure \ref{fig:phase-diagram} a simplified phase diagram of hole-doped cuprates in which the pseudogap, strange metal and FL regimes are framed along the $T$ = 0 doping axis and delineated by vertical dashed lines. In so doing, we define the strange metal regime as the doping range ($p^{\ast} < p < p_{sc}$) over which a dominant low-$T$ $T$-linear resistivity prevails  \cite{cooper_science_2009, hussey_2013, legros_natphys_2019, Putzke_NatPhys_2021}. For simplicity, only a singular boundary line separating each regime has been drawn, though it is anticipated that the precise location of each boundary will be slightly material specific.

The double-headed arrows in Fig.~\ref{fig:phase-diagram} define the doping ranges of the various families investigated in this study. The study itself includes new MR data on LSCO crystals with $x$ = 0.20, 0.23, 0.30 and 0.33, as well as new analysis of MR data reported elsewhere on LSCO ($x$ = 0.09 \cite{Kimura96}, 0.15 \cite{harris_prl_1995}, 0.19 \cite{giraldogallo_science_2018} and 0.23 \cite{cooper_science_2009}). In the case of Bi2201, we report new MR data for $p$ = 0.13, 0.16, 0.195, 0.205, 0.22 and 0.24 and 0.245 (much of it taken in pulsed fields up to 60 T) combined with further analysis of lower-field MR data reported in Ref.~\cite{Ayres_2021} for the restricted doping range 0.24 $< p <$ 0.27. For each family, we have performed field sweeps at multiple temperatures from below $T_c(0)$ up to 300 K. For Tl2201, we report MR data on a single sample whose $T_c$ (and therefore $p$) is varied using hydrostatic pressure and whose ambient pressure data were reported in Ref.~\cite{Ayres_2021}. Also included in Fig.~1 is a single Hg1201 sample reported in Ref.~\cite{chan_prl_2014}.


\begin{figure*}[!ht]
              \centering        
              \includegraphics[width=0.9\textwidth]{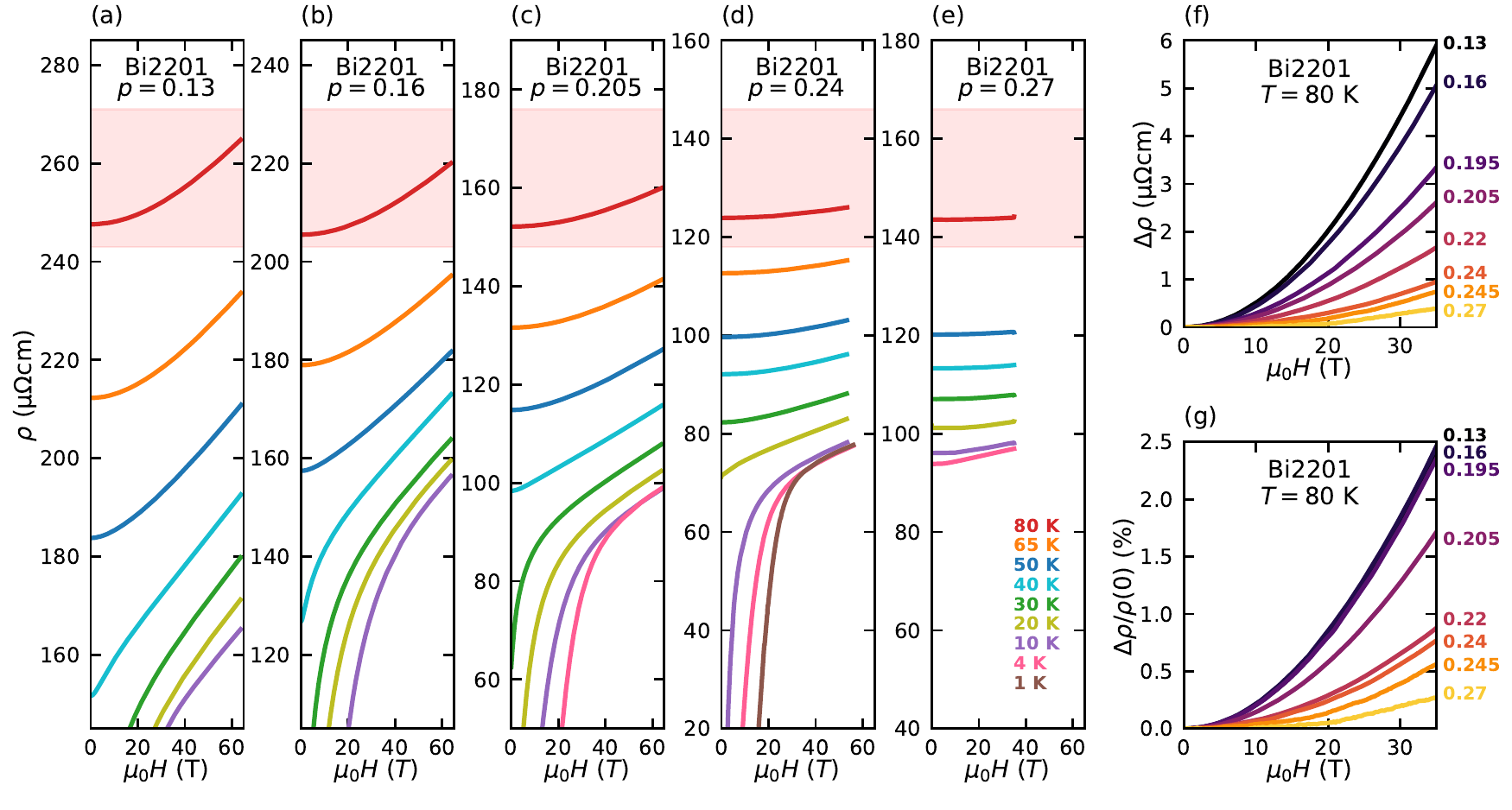}
              \caption{(a)-(e): $\rho(H, T)$ of a selection of Bi2201 crystals spanning the doping range 0.13 $\leq p \leq$ 0.27. With decreasing $p$, a marked increase in the magnitude of the magnetoresistance (at a constant temperature) is observed, as highlighted within the red shaded regions for $T$ = 80 K. (f) $\Delta\rho(H, 80 \mathrm{K})$ curves for all 8 dopings studied. (g) Corresponding $\Delta \rho/\rho(0)$ curves.  For $p > p^\ast \sim 0.2$, the magnitude of $\Delta \rho/\rho(0)$ drops with increasing $p$, revealing that within the strange metal regime ($p^\ast < p < p_{sc}$), the order-of-magnitude enhancement in the MR is not a result of an increase in $\rho(0,T)$ (e.g. due to a reduced carrier density).}
              \label{fig:Bi2201_MR}
\end{figure*}

Representative series of MR curves within the strange metal regime are shown in panels (a)-(c) of Fig.~\ref{fig:data-summary} for Bi2201, Tl2201 and LSCO, respectively. Adherence to $H/T$ scaling is evident through inspection of panels (d)-(f) where plots of $\Delta\rho/T (= (\rho(H,T)-\rho(0,T)/T)$ vs.~$H/T$ collapse onto, or asymptotically approach (for $T < T_c(H$=0)), the dashed lines defined by Eq.~(\ref{eq:QuadT_2}). Panels (g)-(i) of Fig.~\ref{fig:data-summary} show the derivatives d$\rho$/d$H$, and highlight the tendency of all MR curves to become linear at high $H/T$. The collapse of the d$\rho$/d$H$ curves at the highest fields indicates that the LMR slope $\gamma$ is indeed $T$-independent. The dashed lines in the lower panels are fits of the derivative of Eq.~(\ref{eq:QuadT_2}) to the d$\rho$/d$H$ data. The key aspects of the MR: the crossover from $H^2$ to $H$-linear MR with increasing $H/T$, the constancy of $\gamma$ over an extended temperature range and the adherence to the $H/T$ quadrature scaling form, are apparent in all three data sets and demonstrate universal scaling behavior across different families within the strange metal regime.

Having confirmed the robustness of $H/T$ quadrature scaling in this region of the phase diagram \cite{Ayres_2021}, we now turn to address each of the outstanding questions (\textbf{A}-\textbf{D}) listed above. In the process, we reveal that while LMR is a ubiquitous feature of the high-field MR, the scaling itself is distinct in each regime, the reasons for which we will discuss in Section III.

\subsection{Doping dependence of the $H$-linear MR}

In both electron- and hole-doped cuprates, $a_1$ decreases with increasing doping (beyond $p^{\ast}$) and vanishes around $p \sim p_{sc}$ \cite{cooper_science_2009, Jin_Nature_2011, hussey_2013, legros_natphys_2019, yuan_2021}, revealing a robust correlation between $a_1$ and $T_c$ \cite{yuan_2021}. While previous cuprate studies \cite{giraldogallo_science_2018, sarkar_correlation_2019} have hinted at a similar trend in $\gamma$, the doping ranges were too narrow (0.16 $\leq p \leq$ 0.19 \cite{giraldogallo_science_2018} and 0.15 $\leq p \leq 0.17$ \cite{sarkar_correlation_2019}) to make any definitive claims. In this section, we explore the evolution of $\gamma(p)$ across the SC dome and compare it with $a_1(p)$.

\subsubsection{Bi2201}

Panels (a)-(e) of Figure \ref{fig:Bi2201_MR} show $\rho(H,T)$ for a representative set of Bi2201 samples over a doping range (0.13 $\leq p \leq$ 0.27) that incorporates both the pseudogap and strange metal regimes. At temperatures above $T_F$ (the temperature below which paraconductivity contributions from SC fluctuations become appreciable), the low-field MR is quadratic. In highly doped samples with a low $T_c$ (and thus low $T_F$), the crossover from quadratic to LMR is clearly visible within the accessible field range. This is in agreement with the data reported in Ref.~\cite{Ayres_2021}.
 
The red shaded regions of panels (a)-(e) highlight the doping evolution of the MR at a fixed temperature $T$ = 80 K, i.e.~well above $T_F$ for each sample. Noting that the absolute span in $\rho$ is the same (= 140 $\mu\Omega$cm) in all panels, we find that the magnitude of the MR decreases monotonically with increasing $p$. The overall trend is summarized in Fig.~\ref{fig:Bi2201_MR}f where $\Delta\rho(H$, 80 K) ($=\rho(H$, 80 K) - $\rho$(0, 80 K)) is plotted for all 8 dopings studied. $\rho$(0, 80 K) itself decreases roughly by a factor of 2 across the series, reflecting changes in the scattering rate, the carrier density or some combination of the two. $\Delta\rho(H$, 80 K), by contrast, decreases by more than one order of magnitude. The corresponding $\Delta \rho(H$, 80K)/$\rho$(0, 80 K) values are plotted in Fig.~\ref{fig:Bi2201_MR}g. Intriguingly, for the three pseudogapped samples with $p < p^{\ast} (\sim 0.2$), $\Delta \rho(H$, 80K)/$\rho$(0, 80 K) is found to be unchanged, implying that the increase in the magnitude of $\Delta\rho$ for these samples is a direct consequence of the increase in $\rho(0)$ with underdoping, presumably due to the loss of electronic states once the pseudogap opens. 

For higher dopings, by contrast, even the fractional change in MR is found to vary strongly with $p$, decreasing by an order of magnitude between $p$ = 0.205 and 0.27 (see Fig.~\ref{fig:Bi2201_MR}g). In a typical metal, the magnitude of $\Delta\rho/\rho(0)$ is set by the product $\omega_c\tau$ where $\omega_c = e \mu_0 H/m^*$ is the cyclotron frequency, $e$ is the electronic charge and $m^*$ is the cyclotron mass and is thus largely independent of carrier density. A monotonic decrease in $\tau$ as the system is doped away from the Mott insulating state (and thus becomes increasingly more metallic) seems unlikely; there is little variation in $\rho_0$ within this doping range and no indication, e.g. from existing specific heat measurements \cite{girod-thesis}, for a large change in $m^*$. Thus, the marked decrease in $\Delta\rho/\rho(0)$ with overdoping seems puzzling, though may arise from a decrease in the anisotropy of the in-plane mean-free-path $\ell$ as $p$ approaches $p_{sc}$. Note that any decrease in the anisotropy of $\ell$ cannot be caused by the Fermi energy $\epsilon_F$ moving away from a van Hove singularity (vHs) since in Bi2201, the vHs crossing point is believed to be located closer to $p_{sc} \sim 0.27$ than to $p^{\ast} \sim 0.2$ \cite{Ding_2019}.
 
 \begin{figure}[!ht]
              \centering        
              \includegraphics[width=0.45\textwidth]{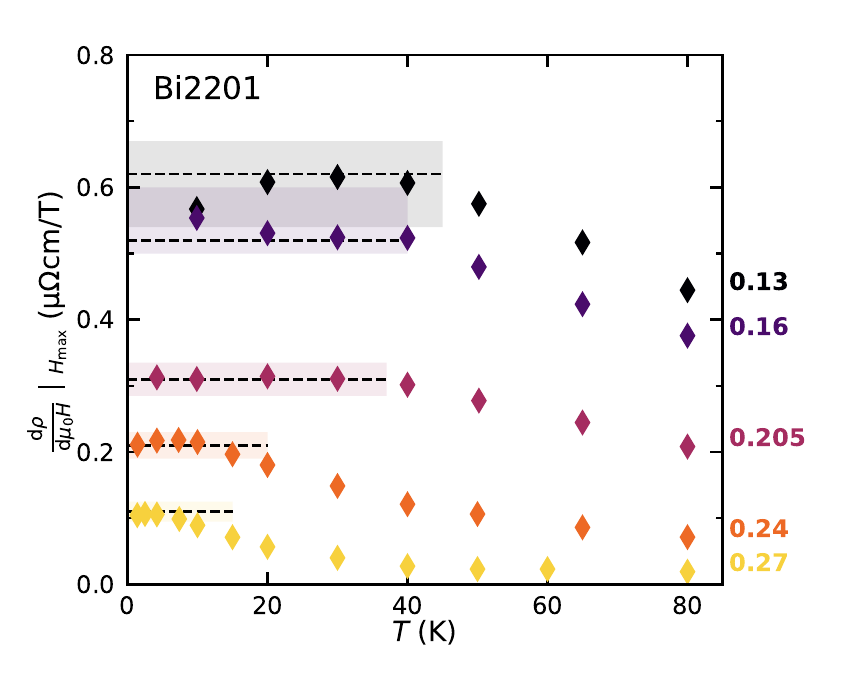}
              \caption{$T$-dependence of the  derivative d$\rho$/d$\mu_0 H$ at the highest measured field range (specifically the last 3 T of each field sweep) -- for the MR curves displayed in panels (a)-(e) of Fig.~\ref{fig:Bi2201_MR}. For all dopings, d$\rho$/d$\mu_0 H|_{H_{\rm max}}$ saturates at low $T$ at a value (= $\gamma$) denoted by horizontal dashed lines. Shaded regions reflect the uncertainty in $\gamma$ for each data set. Estimates for $\gamma$ obtained from field sweeps for which the $H$-linear regime could not be reached are not shown. For $p$ = 0.16, the upturn in d$\rho$/d$\mu_0 H|_{H_{\rm max}}$ at the lowest $T$ is due to paraconductivity effects. The origin of the low-$T$ downturn in d$\rho$/d$\mu_0 H|_{H_{\rm max}}$ for $p$ = 0.13 is unknown. Nevertheless, an estimate for $\gamma$ can still be made, albeit with greater uncertainty.}
              \label{fig:gamma-vs-temperature}
\end{figure}

At high $H/T$ (below $T_c$ in most samples) and for all dopings, the MR becomes $H$-linear with a $T$-independent slope $\gamma$, consistent with the quadrature expression in Eq.~(\ref{eq:QuadT_2}). Such behavior can also be inferred from the high-$H$ convergence of the d$\rho$/d$H$ curves plotted in Fig.~\ref{fig:data-summary}. In order to obtain estimates for $\gamma(p)$, we study the low-$T$ limit of the highest field data (where the normal state can be accessed). A summary of the results is shown in Figure \ref{fig:gamma-vs-temperature} where the $T$-dependence of d$\rho$/d$\mu_0 H|_{H_{\rm max}}$ -- the MR slope at the highest measured field range (i.e.~over the last 3 T of each field sweep) -- is plotted for all the samples whose raw MR data are included in panels (a)-(e) of Fig.~\ref{fig:Bi2201_MR}. In agreement with Eq.~(\ref{eq:QuadT_2}) and with previous measurements on LSCO \cite{giraldogallo_science_2018}, Tl2201 and Bi2201 \cite{Ayres_2021}, d$\rho$/d$\mu_0 H|_{H_{\rm max}}$ is found to increase with decreasing temperature and to saturate at a constant value (= $\gamma$) at the lowest temperatures. Consistent with our expectations from Fig.~\ref{fig:Bi2201_MR}g, we observe a marked monotonic increase in $\gamma(p)$ as $p$ is reduced.

\subsubsection{Tl2201}

Although the increase in $\gamma$ with decreasing $p$ shown in Fig.~\ref{fig:gamma-vs-temperature} is significant, the above results involve measurements on different samples. Inevitably, inter-sample variation and geometrical errors will introduce some uncertainty in the determination and evolution of $\gamma(p)$. In order to circumvent this, a second study was carried out on a single crystal of overdoped Tl2201 ($T_c$ = 35 K at ambient pressure) as a function of hydrostatic pressure. The $T_c$ (and inferred doping) of Tl2201 can be tuned appreciably with the application of only modest pressures \cite{looney_2001}. Panels (a)-(e) of Figure 2 in the Supplementary Information (SI) show the full set of $\rho(H,T)$ curves obtained across the pressure range 0 to 2 GPa, over which $T_c$ decreases from 35 K to 26 K (26 \%). Panels (f)-(i) of the same figure show the corresponding derivatives plotted vs $H/T$. A gradual decrease in $\gamma$ is observed with increasing pressure, consistent with the correlation between $\gamma$ and $T_c$ (or $p$) seen in Bi2201 beyond $p^\ast$.

\begin{figure*}[!ht]
              \centering        
              \includegraphics{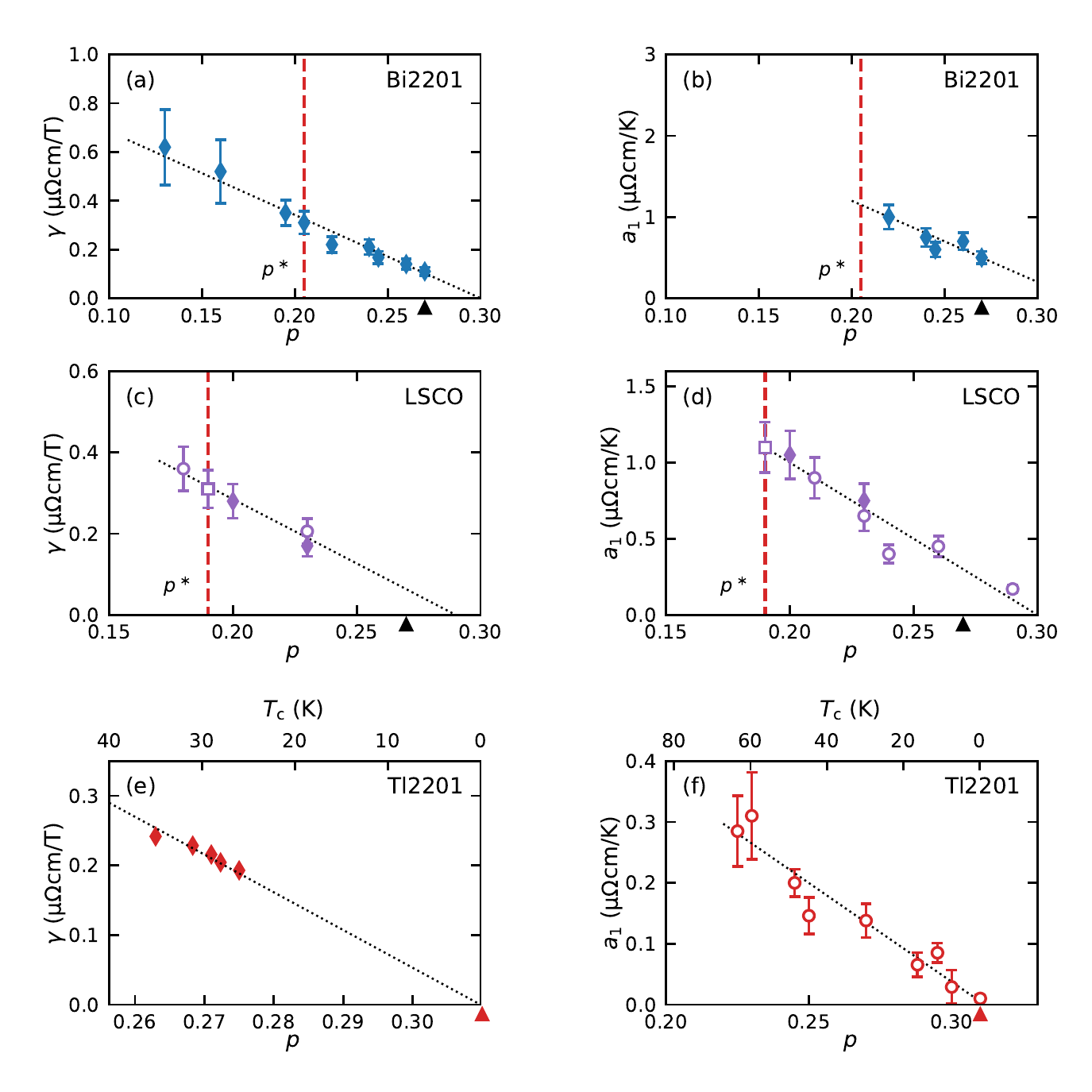}
              \caption{Doping dependence of $\gamma$ -- the high $H/T$ limiting value of the MR -- and $a_1$ -- the low-$T$ $T$-linear coefficient of the resistivity -- for (a,b) Bi2201, (c,d) LSCO and (e,f) Tl2201, respectively. Diamonds indicate results obtained as part of this high-field study, while circles and squares indicate results obtained from the literature. Circles and squares in panels (c) and (d) are taken from Ref.~\cite{cooper_science_2009} and \cite{giraldogallo_science_2018}, respectively. Circles in panel (f) are from Ref.~\cite{hussey_2013}. Note that $a_1$ is plotted only for samples with $p > p^\ast$. In panel (e), $\gamma$ is plotted vs.~$p$ as determined from the $T_c$ value at each pressure using the correlation reported in Ref.~\cite{Putzke_NatPhys_2021}. All dotted lines are guides to the eye. Arrowheads on the doping axes indicate $p_{sc}$ for the respective material.}
              \label{fig:gamma-A-vs-doping}
\end{figure*}

\subsubsection{LSCO}

As reported in Ref.~\cite{giraldogallo_science_2018}, thin-film LSCO at a doping level ($x$ = 0.19) near $p^{\ast}$ also exhibits LMR at the highest fields. In order to confirm this behavior and to extend the measurements of the MR up to higher dopings (and higher temperatures), we measured the MR of two superconducting LSCO single crystals with $p$ = 0.20 and 0.23 up to 35 T. The derivatives $\mathrm{d}\rho/\mathrm{d}H$ show that, just as in overdoped Bi2201 and Tl2201, the MR becomes $H$-linear at high $H/T$ values (the data for $x = p$ = 0.23 are presented in Fig.~\ref{fig:data-summary}). In an earlier study \cite{cooper_science_2009}, high fields were employed to obtain the limiting low-$T$ resistivity in overdoped LSCO over a wide doping range, though there, the normal state MR was assumed to vary as $H^2$ at all field strengths up to 55 T. Re-analysis of their MR data for $x$ = 0.23, presented in Fig.~3 of the SI, combined with the new measurements, confirms that the high-field MR in overdoped LSCO also shows a tendency towards LMR at the lowest temperatures.

\subsubsection{Summary}

The left and right panels of Figure \ref{fig:gamma-A-vs-doping} show, respectively, the doping/pressure evolution of $\gamma$ and $a_1$ for all three families. Although an equivalence between pressure and hole doping in Tl2201 has not yet been definitively demonstrated, we adopt here the correlation between $T_c$ and $p$ reported in Ref.~\cite{Putzke_NatPhys_2021} to arrive at the same conclusion. The striking correlation between $\gamma$ and $a_1$, combined with the ubiquitous $H/T$ scaling beyond $p^{\ast}$, confirms that the two phenomena share a common origin (though why the ratio $\gamma/a_1$ is higher in Tl2201 than in LSCO or Bi2201 is not yet clear). Crucially, $\gamma(p)$ appears to extrapolate to zero at a doping level $p \approx$ 0.3 for all three families, implying that the emergence of LMR and $H/T$ scaling in hole-doped cuprates is closely tied to the onset of superconductivity, as recently intimated for electron-doped cuprates \cite{sarkar_correlation_2019}.  

\begin{figure*}[!ht]
              \centering        
              \includegraphics{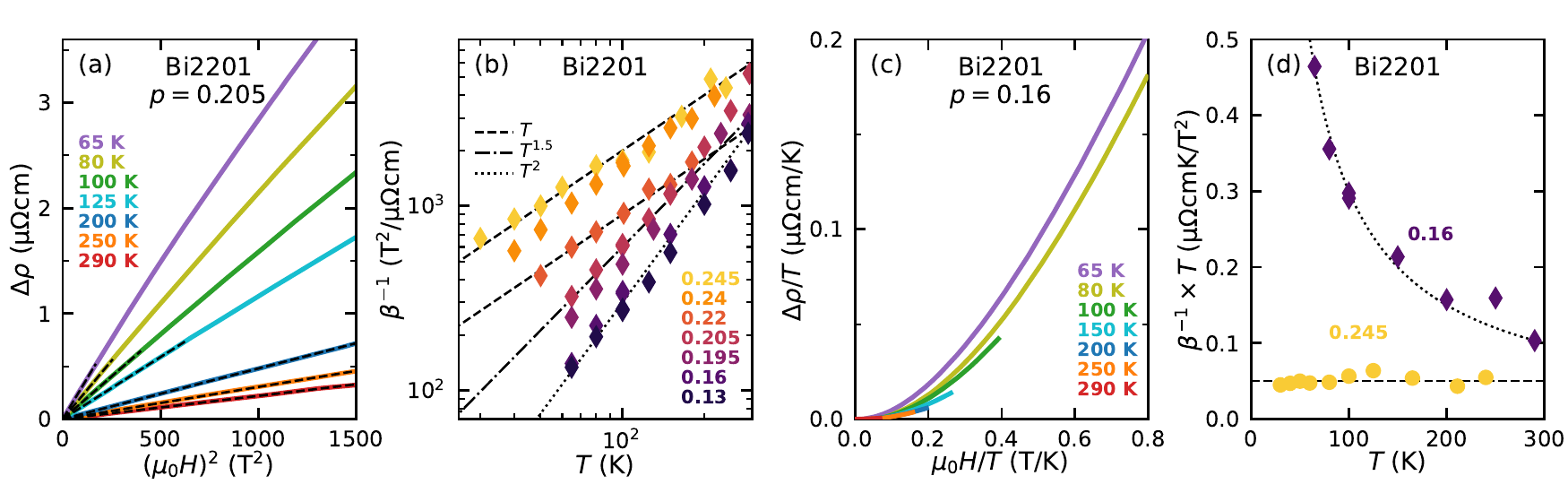}
              \caption{(a) $\Delta\rho (= \rho(T,H) - \rho(T,0)$) plotted versus $H^2$ in Bi2201 $p$ = 0.205 highlighting the $H^2$ dependence of the MR at low fields. The dashed lines are quadratic fits ($\Delta\rho = \beta(\mu_0 H)^2$) to the data in the low $H/T$ range from which the $\beta$ coefficients are obtained. (b) Resultant $\beta^{-1}$ values plotted versus temperature on log-log axes. The dashed, dash-dotted and dotted guides are shown for $m$ = 1, 1.5 and 2, respectively. (c) $\Delta\rho/T$ vs.~$H/T$ plot for optimally doped Bi2201 ($p$ = 0.16). Clearly, the data do not collapse. (d) $\beta^{-1} \times T$ vs.~$T$ for Bi2201 $p$ = 0.16 and 0.245. The dotted line represents a 1/$T$ dependence.}
              \label{fig:Bi2201-scaling-demonstration}
\end{figure*}

This correlation between $\gamma$ and $a_1$ within the strange metal regime is robust and seemingly universal, having been observed in three different families. Its preservation below $p^{\ast}$, however, is far less certain since the low-$T$ $T$-linear component in $\rho(0,T)$ is either reduced \cite{cooper_science_2009, giraldogallo_science_2018}, or lost entirely \cite{Rullier_Albenque08, chan_prl_2014, proust_PNAS_2016, Berben_PD_2022}. The observation of a continued increase in $\gamma$ below $p^{\ast}$ is thus surprising as it suggests an apparent decoupling of $\gamma$ and $a_1$ below $p^{\ast}$, an anomaly we will return to in the Discussion section. But first, having established $H/T$ scaling as a universal property of the strange metal regime, let us turn to examine how the MR scaling itself evolves as the pseudogap develops.


\subsection{MR scaling within the pseudogap regime}

As the pseudogap opens, electronic states near the Fermi level become depleted, initially near the zone boundary. At the same time, the $T$-linear component in $\rho(0,T)$ (below $T^{\ast}$) becomes suppressed and $\rho(0,T)$ recovers a dominant $T^2$ dependence \cite{proust_PNAS_2016, chan_prl_2014}. In the clean cuprate Hg1201, the MR is also reported to exhibit more conventional, Kohler-like scaling \cite{chan_prl_2014}. This, coupled with the observation of quantum oscillations, has led to the conjecture that the remnant charge at the Fermi level in the low-$T$ pseudogap regime forms a Fermi liquid. In order to differentiate between Kohler and quadrature scaling, however, it is helpful to investigate the MR response of underdoped cuprates with larger residual resistivities.

Until now, $H/T$ scaling has been revealed by collapsing plots of $\Delta\rho/T$ or d$\rho/$d$H$ vs.~$H/T$ onto a single curve. For certain field sweeps, however, the maximum field strength is not high enough to confirm the scaling relation. An alternative way to test for such scaling is to focus on the low-field ($H \to 0$) regime where $\Delta\rho(H)$ asymptotically approaches a $H^2$-dependence. Rewriting Eq.~(\ref{eq:QuadT_2}) as:
\begin{eqnarray}
\Delta\rho (T, H) & = & \alpha T \left( \sqrt{1 + \left(\gamma \mu_0 H/\alpha T\right)^2} -1\right)
\label{eq:QuadT_1}
\end{eqnarray}
and performing a Taylor expansion, we find:
\begin{eqnarray}
\Delta\rho (T, H \rightarrow 0) & = & \frac{\gamma^2}{2\alpha T} (\mu_0 H)^2 = \beta (\mu_0 H)^2
\label{eq:B_term}
\end{eqnarray}
Hence, for any material exhibiting quadrature MR with pure $H/T$ scaling, the coefficient $\beta$ of the low-field $H^2$ MR should vary as $1/T$. More generally, a quadrature form of the MR scaling as $H/T^m$ (as opposed to the $H/T$ scaling form expressed through Eq.~(\eqref{eq:QuadT_1})) will lead to $\beta \propto 1/T^m$. This is in marked contrast to the $\beta \propto$ 1/$\rho(0,T)$ dependence expected from Kohler scaling (recall that in LSCO and Bi2201, $\rho_0$ is large relative to the $T$-dependent part of $\rho(0,T)$ -- see, e.g.~Fig.~2). With this in mind, we now turn to examine the $T$-dependence of $\beta^{-1}$ as we approach $p^\ast$ and as the pseudogap opens (see Fig.~S4 and associated text in the SI for a discussion of the robustness of the extraction of $\beta(T)$ through this procedure). Fig.~\ref{fig:Bi2201-scaling-demonstration}a illustrates how $\beta(T)$ is determined for a Bi2201 sample with $p$ = 0.205 (close to $p^\ast$), while Fig.~\ref{fig:Bi2201-scaling-demonstration}b shows resultant plots of $\beta^{-1}(T)$ for all Bi2201 samples on a log-log scale.

The sequence of straight lines in Fig.~\ref{fig:Bi2201-scaling-demonstration}b reveals two important features of the low-field MR of Bi2201. Firstly, $\beta^{-1}(T)$ exhibits a pure power-law $T$-dependence at all dopings and for all $T > T_{\mathrm{F}}$ at which the normal state MR can be accessed down to $H$ = 0. Secondly, as indicated by the dashed/dotted lines, the exponent $m$ of the power-law dependence is found to change monotonically from 1 to 2 (see legend) as doping is reduced. It is important to reiterate that the high residual resistivities of Bi2201 are such that, were Kohler scaling to be obeyed, marked deviations from such power-law fits would be evident at temperatures far higher than the base temperature of these measurements. This particular feature of the data will become even more evident in Section II.C when we consider highly overdoped, non-SC LSCO.

\begin{figure*}[!ht]
              \centering        
              \includegraphics{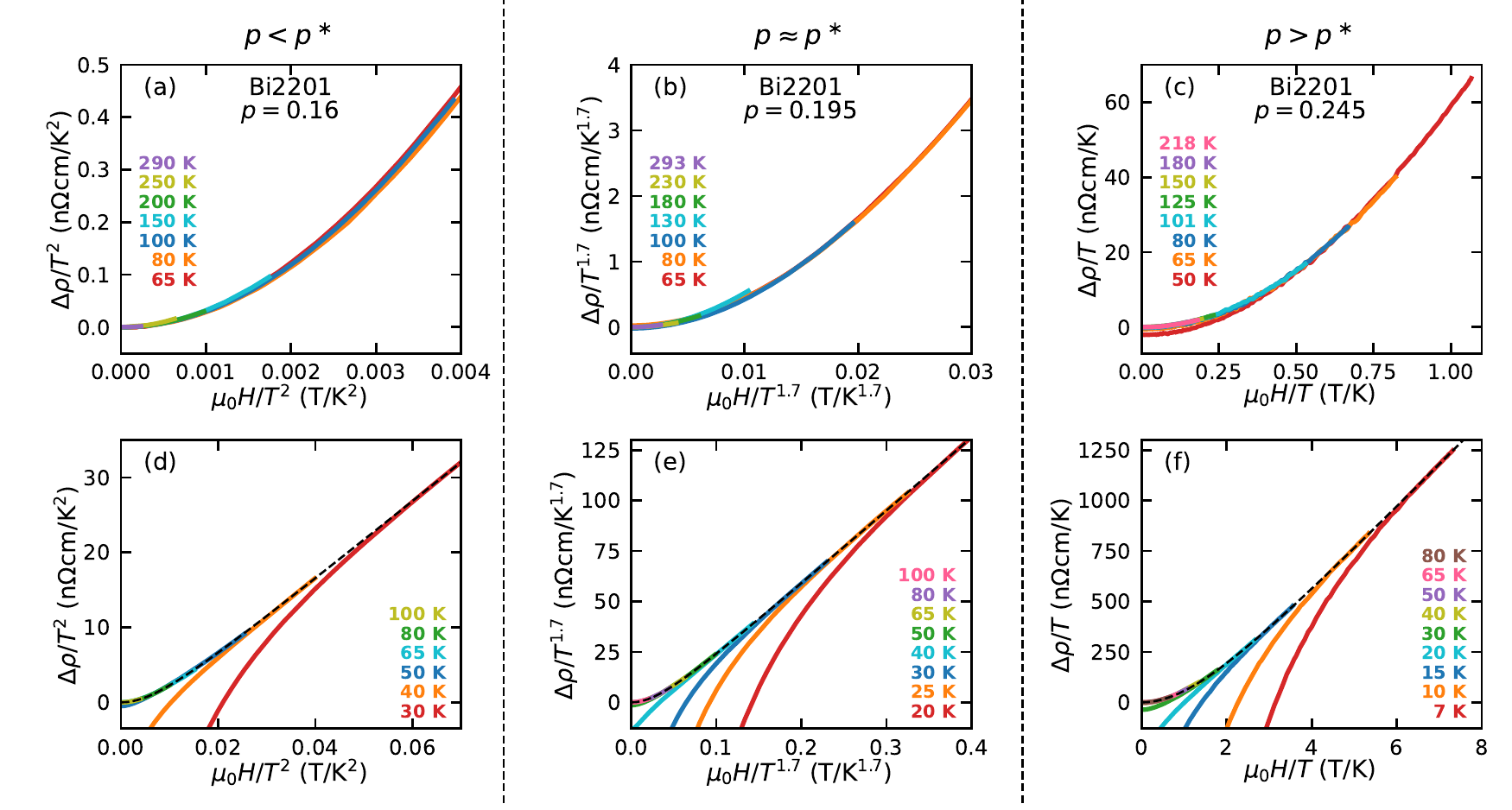}
              \caption{Scaling plots for selected Bi2201 samples where the appropriate exponent $m$ for each sample is used to rescale both axes with $T^m$ and collapse the MR data. (a) $p$ = 0.16: $\Delta\rho/T^2$ plotted versus $\mu_0 H/T^2$.  (b) $p$ = 0.195: $\Delta\rho/T^{1.7}$ plotted versus $\mu_0 H/T^{1.7}$. (c) $p$ = 0.245: $\Delta\rho/T$ plotted versus $\mu_0 H/T$. For panels (a)-(c), only the MR above $T_{\mathrm{F}}$ is plotted. (d)-(f): Same scaling plots as in panels (a)-(c) for temperatures below $T_{\mathrm{F}}$ and over an extended $H/T^m$ range. The black dashed lines indicate the quadrature form obtained from fits to the normal-state data using Eq.~(\ref{eq:QuadT2_1}).}
              \label{fig:Bi2201-power-law-evolution}
\end{figure*}

The crossover from $H/T$ ($m=1$) scaling within the strange metal regime to $H/T^2$ ($m=2$) scaling below $p^\ast$ is further highlighted in panels (c) and (d) of Figure \ref{fig:Bi2201-scaling-demonstration}. Fig.~\ref{fig:Bi2201-scaling-demonstration}c captures the loss of $H/T$ scaling for $p = 0.16$ (compare Fig.~\ref{fig:Bi2201-scaling-demonstration}c with panels (c) and (f) of Fig.~\ref{fig:Bi2201-power-law-evolution} for a Bi2201 sample with $p$ = 0.245, i.e. deep within the strange metal regime). Fig.~\ref{fig:Bi2201-scaling-demonstration}d shows the product $T\beta^{-1}$ plotted vs.~$T$ for the same two samples. For $p$ = 0.245, $T\beta^{-1}$ is a constant, consistent with $m$ = 1, while for $p$ = 0.16, $T\beta^{-1}$ diverges (as $1/T$) as $T$ decreases.

Despite these marked changes in the power-law exponent of the low-field MR, the field-dependence of the MR evolves in all cases from quadratic at low fields to linear at high fields with a linear slope that saturates at high fields (and/or low temperatures). This behavior can be captured by the following generalized expression:

\begin{eqnarray}
\Delta\rho(H,T) & = & \alpha_m T^m \left( \sqrt{1 + \frac{(\gamma \mu_0 H)^2}{(\alpha_m T^m)^2}} -1\right)~,
\label{eq:QuadT2_1}
\end{eqnarray}
where $\alpha_m$ is a prefactor independent of $T$ and $H$ with units $\mu\Omega$cm/K$^m$. The total resistivity is then expressed as:
\begin{eqnarray}
\rho(H, T) & = & \mathcal{F}(T)+\sqrt{(\alpha_m T^m)^2 + (\gamma \mu_0 H)^2}~.
\label{eq:QuadT2_2}
\end{eqnarray}

To illustrate the goodness of fit to Eq.~(\ref{eq:QuadT2_1}), Figure \ref{fig:Bi2201-power-law-evolution} shows MR curves for three representative Bi2201 samples, where the appropriate exponent $m$ for each sample is used to rescale both axes by $T^m$ and collapse the MR data. Note that the data collapse occurs over a decade in temperature in all three cases. Intriguingly, while $m$ has an integer value in the pseudogap and strange metal regimes, in the vicinity of $p^\ast$, $m$ takes on a value that is intermediate between the two.

\begin{figure}[!ht]
              \centering        
              \includegraphics{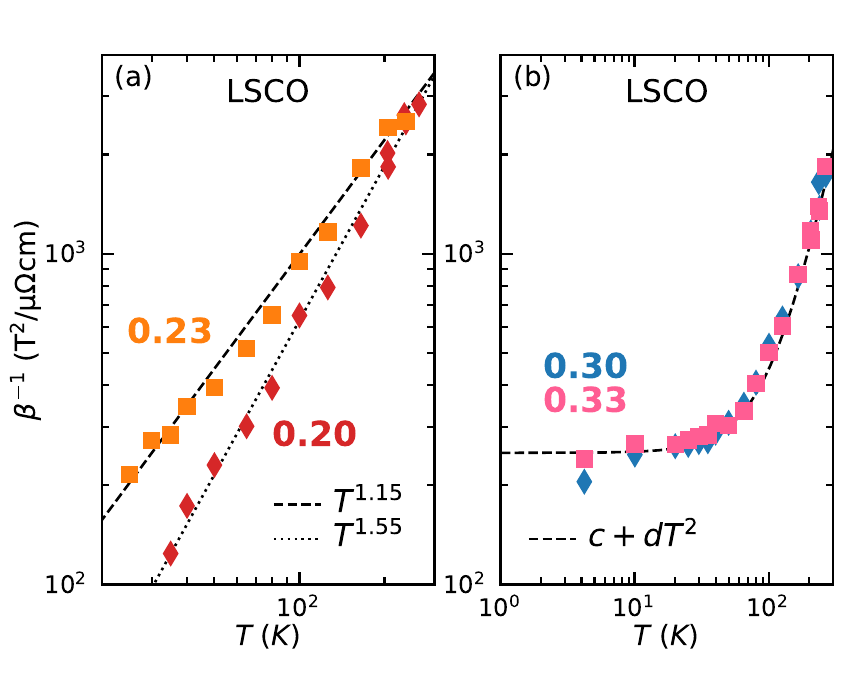}
              \caption{(a) Log-log plot of $\beta^{-1}(T)$ for LSCO near $p^\ast$ ($x = 0.20$, filled diamonds) and beyond ($x = 0.23$, filled squares). Dashed and dotted lines represent power laws $T^{1.15}$ and $T^{1.55}$, respectively. (b) Comparative plot for LSCO dopings beyond the SC dome ($x = 0.30$, diamonds; $x = 0.33$, squares). The dashed line indicates that $\beta^{-1}(T) \sim c + dT^2$.}
              \label{fig:lsco-scaling}
\end{figure}

Having established the ubiquity of power-law scaling in superconducting Bi2201, we turn to investigate whether similar behavior is realized in other cuprate families. In Figure \ref{fig:lsco-scaling}a, we plot $\beta^{-1}$ vs.~$T$ for two superconducting LSCO samples, one close to $p^{\ast}$ ($x$ = 0.20) and one inside the strange metal regime ($x$ = 0.23). For $x$ = 0.23, a pure power-law dependence is observed from the highest temperature range studied down to 25 K with  $m = 1.15\pm0.1$, in accord with the collapse of the $\Delta\rho/T$ vs.~$H/T$ curves displayed in Fig.~\ref{fig:data-summary}. For $x$ = 0.20, $\beta^{-1}(T)$ also follows a pure power-law dependence, though in this case, $m = 1.55\pm 0.15$, a value that is similar to those found in Bi2201 in the vicinity of $p^{\ast}$. Previously, it was shown that the MR data for LSCO $x$ = 0.19 \cite{giraldogallo_science_2018} do not exhibit $H/T$ quadrature scaling \cite{boyd_prb_2019} even though LMR is observed at high field strengths. Our own analysis of the data of Ref.~\cite{giraldogallo_science_2018} -- shown in Figure 5 of the SI -- reveals that the MR for $x$ = 0.19 actually exhibits $H/T^2$ scaling. Fig.~6 of the SI also shows that for optimally doped LSCO ($x$ = 0.17 \cite{harris_prl_1995}) and underdoped Hg1201 ($p = 0.10$ \cite{chan_prl_2014}), $\beta^{-1} \propto T^2$. This latter finding is consistent with the original conclusions of Chan \textit{et al.} \cite{chan_prl_2014}, who claimed that the MR of underdoped Hg1201 obeys conventional Kohler scaling. Being one of the cleanest cuprate systems, $\rho_0$ of underdoped Hg1201 is extremely low and since $\rho(0,T)$ $\propto$ $T^2$ below around 200 K \cite{chan_prl_2014}, a Kohler plot (with $\Delta\rho/\rho(0)$ vs.~$H/\rho(0)$) does indeed produce the same scaling. As a result, both descriptions are justified. In underdoped LSCO or Bi2201, on the other hand, Kohler scaling clearly breaks down due to the fact that their residual resistivities are much larger. Thus, we conclude that $H/T^2$ scaling is the more appropriate way to categorize the universal low-field MR response of hole-doped cuprates below $p^\ast$ and that the absence of $\rho_0$ in the MR scaling is a notable feature which cannot be captured by Boltzmann theory, highlighting a distinctly non-FL aspect to the MR within the pseudogap regime.


\subsection{Scaling of the MR beyond the SC dome}

As stated in the Introduction, for $p > p_{sc}$, $\rho(0,T) \propto T^2$ at low-$T$, as expected for a correlated FL \cite{Manako_PRB_1992, nakamae_2003}. Moreover, in electron-doped cuprates, the field-dependence of the low-field MR switches from linear to quadratic as superconductivity is suppressed \cite{sarkar_correlation_2019}. The form of the MR in hole-doped cuprates beyond $p_{sc}$, however, has not yet been investigated. Here, we report a recovery of Kohler scaling in $p$-doped cuprates beyond the SC dome.

\begin{figure}[!ht]
        \centering        
        \includegraphics{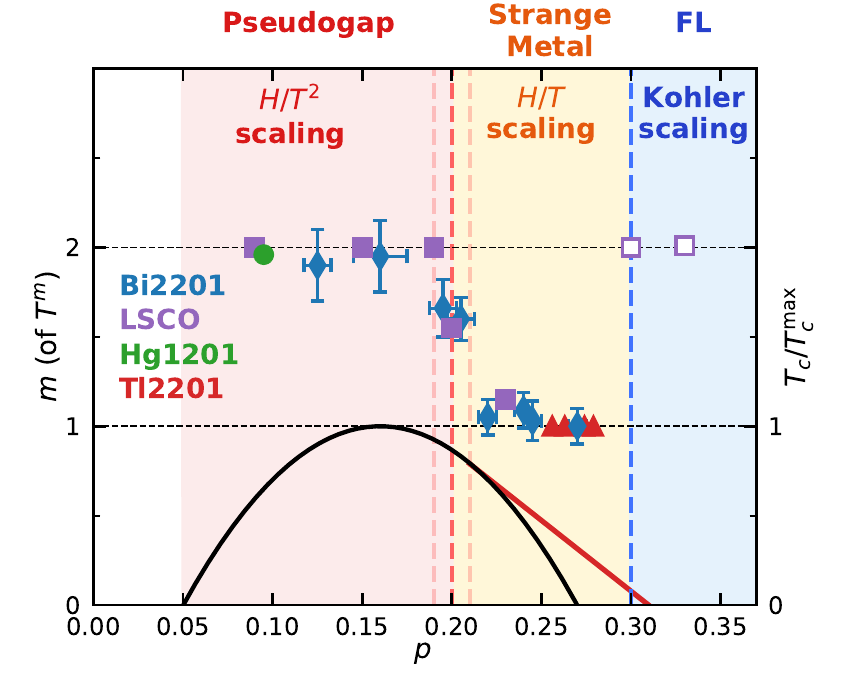}
        \caption{Doping dependence of the power-law MR scaling in hole-doped cuprates. Right axis: Normalized $T_c/T_c^{\rm max}$ for LSCO and Bi2201 (black line) and Tl2201 (red line). Left axis: The exponent $m$ as determined from a power law fit to $\beta^{-1}$ ($\sim T^m$) for Bi2201, LSCO and Hg1201. This plot combines data obtained in this work with published data on LSCO \cite{Kimura96, harris_prl_1995, giraldogallo_science_2018} and Hg1201 \cite{chan_prl_2014}. As the temperature range to extract $\beta^{-1}(T)$ from the low-field MR in Tl2201 is too narrow, the data points for Tl2201 were deduced from the $H/T$ scaling first reported in Ref.~\cite{Ayres_2021} and confirmed in Fig.~S2 of the SI. As the pseudogap closes, $m$ transitions rapidly from 2 to 1. In the FL regime beyond $p_{sc}$, conventional Kohler scaling (with $\beta^{-1} \sim c + dT^2$) is restored (open squares).}
        \label{fig:exponent}
\end{figure}

Figure \ref{fig:lsco-scaling}b shows log-log plots of $\beta^{-1}(T)$ for two highly-doped non-SC LSCO crystals. It is immediately evident that in contrast to the SC samples, $\beta^{-1}(T)$ does not follow a simple power-law, but instead tends towards a constant value as $T \to$ 0. (Note that the curvature in $\beta^{-1}(T)$ (on a log-log scale) sets in at a temperature $T \sim$ 80 K that is significantly higher than the lowest temperatures at which pure power-law scaling is found to persist in SC samples.) As indicated by the dashed line in Fig.~\ref{fig:lsco-scaling}b, the data can be fitted well to the expression $\beta^{-1}(T) = c+dT^2$, consistent with the form of $\rho(0,T)$ at low $T$ (see Figs.~S6 and S7 of the SI) and thus with more conventional Kohler scaling. 

The observed recovery of a finite residual component (= $c$ in Fig.~\ref{fig:lsco-scaling}b) in the MR of non-SC samples reveals a third regime of distinct MR behavior in hole-doped cuprates. The overall picture is summarized in Figure \ref{fig:exponent} where the exponent $m$ (of $\beta^{-1}(T)$) is plotted versus $p$ for all three families studied here together with the earlier result on Hg1201 \cite{chan_prl_2014}. The most striking feature of Fig.~\ref{fig:exponent} is the step-wise change in the exponent from $m \approx$ 2 throughout the pseudogap regime, to $m \approx$ 1 within the strange metal regime over a crossover region $\Delta p$ that is much narrower than any of the regimes themselves. The sharpness of the crossover is reminiscent of the transition from coherent to incoherent anti-nodal ARPES spectra across $p^{\ast}$ first observed in bilayer Bi2212 \cite{chen_incoherent_2019} and later confirmed in Bi2201 \cite{Berben_PD_2022}. This summary plot is the central finding of this work whose significance we shall discuss in more detail in Section III.

\begin{figure*}[!ht]
              \centering        
              \includegraphics{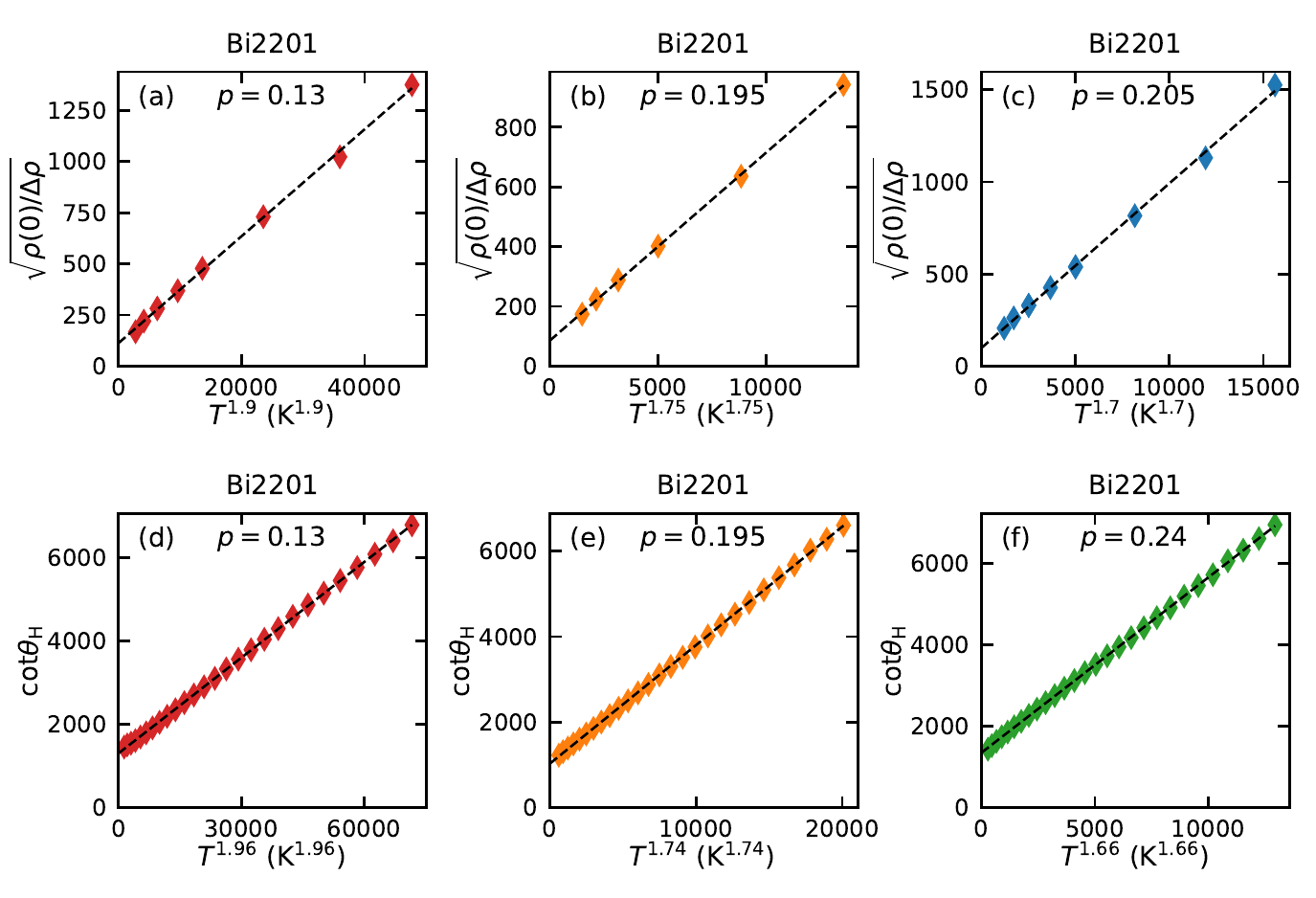}
              \caption{Comparison of the $T$- and $p$-dependences of the normal state MR and Hall angle in Bi2201. Top panels: Plots of $\sqrt{\rho(0)/\Delta\rho}$ vs.~$T^n$ for (a) $p$ = 0.13, (b) 0.195 and (c) 0.205. Bottom panels: Plots of cot$\theta_{\rm H}$ vs.~$T^n$ for (d) $p$ = 0.13, (e) 0.195 and (f) 0.24. Note the evolution of $n$ with $p$. The Hall and MR data were obtained on crystals from the same batches. }
              \label{fig:link-to-kohler}
\end{figure*}


\subsection{Link to modified Kohler scaling}

Finally, we turn to address the last of the four outstanding questions; namely, the link between the power-law MR scaling revealed here and modified Kohler scaling. Harris {\it et al.} were the first to show that the $T$-dependence of $\Delta\rho/\rho(0)$ in optimally doped and underdoped cuprates at low fields scales as tan$^2\theta_{\rm H}$ \cite{harris_prl_1995}, behavior that has often been interpreted as evidence for a separation of transport lifetimes \cite{Anderson_PRL_1991}. A similar separation of lifetimes has been reported in Bi2201 although here, cot$\theta_{\rm H} = A + BT^n$ with $n$ varying from $\approx$ 2 below $p^\ast$ to $\approx 1.6$ at higher dopings \cite{Ando_1999, Konstantinovic_1999}. 

Figure \ref{fig:link-to-kohler} shows plots of $\sqrt{\rho(0)/\Delta\rho}$ and cot$\theta_{\rm H}$ vs.~$T^n$ for three different Bi2201 samples with comparable $p$ values: $p$ = 0.13 (panels (a) and (d)), 0.195 (panels (b) and (e)) and 0.205-0.24 (panels (c) and (f)). In each case, $n$ is chosen to generate a straight line. When plotted in this way, we find that $\sqrt{\rho(0)/\Delta\rho}$ can be equally expressed as $A + BT^n$ with $n$ undergoing a similar evolution with doping as cot$\theta_{\rm H}(T)$. Hence, modified Kohler scaling is also observed in Bi2201. Crucially however, the residual term in the MR plots (panels (a)-(c)) has only appeared by dividing $\Delta\rho$ -- that of itself exhibits pure $1/T^n$ scaling -- by $\rho(0, T)$ that contains a large $\rho_0$. In a similar vein, the cot$\theta_{\rm H}(T)$ plots are ratios of $\rho(0, T)$ and another quantity (the Hall coefficient $R_{\rm H}(T)$ in this case). Note that the prefactors $A$ and $B$ in both sets of plots are not identical, reflecting the fact that $R_{\rm H}(T)$ in Bi2201 does not exhibit a pure $1/T$ dependence, at least at low $T$ \cite{Ando_1999, Konstantinovic_1999, Putzke_NatPhys_2021}. The key point, however, is that while modified Kohler scaling is approximately realized in Bi2201, plotting the data in this manner obscures the intrinsic power-law $T$-dependence of $\Delta\rho(H,T)$ and its associated $H/T^m$ scaling. It will be interesting to explore, in due course, whether the modified Kohler scaling reported for YBCO \cite{harris_prl_1995} is also an artefact of their fitting protocol.


\section{Discussion}

This comprehensive study has uncovered a number of new and surprising relations in the normal state magnetotransport of hole-doped cuprates. The MR in Bi2201, LSCO and Tl2201 exhibits $H/T$ scaling (Eq.~\ref{eq:QuadT_2})) throughout the strange metal regime ($p^{\ast}<p<p_{sc}$) with a linear slope $\gamma$ that correlates with the growth of $a_1$ -- the low-$T$ $T$-linear component of $\rho(0,T)$. Moreover, extrapolation of $\gamma(p)$ or $\gamma(T_c)$ in all three families suggests a close correlation between the magnitude of the $H$-linear MR at high $H/T$ and superconductivity. $\gamma$ continues to grow below $p^{\ast}$ while the relative MR $\Delta\rho(H)/\rho(0)$ is found to have the same magnitude at all $p < p^{\ast}$, suggesting that the continued growth in $\gamma$ with decreasing $p$ reflects the loss of carriers inside the pseudogap regime. At the same time, the power-law scaling of $\Delta\rho(H,T)$ transforms in the vicinity of $p^{\ast}$ from $H/T$ to $H/T^2$. Only beyond the SC dome is conventional Kohler scaling restored. Finally, despite the fact that $\sqrt{\rho(0)/\Delta\rho}$ obeys modified Kohler scaling ($\propto$ cot$\theta_{\rm H}(T)$), it is the power-law scaling of $\Delta\rho(H,T)$ that is found to be the most natural and robust way to interpret the MR. In this section, we explore possible origins of the observed phenomenology.

\subsection{Origin of $H$-linear MR in $p$-doped cuprates} 

LMR and $H/T$ scaling have now been reported in several strange and/or quantum critical metals \cite{Hayes_NatPhys_2016, sarkar_correlation_2019, licciardello_coexistence_2019, nakajima_2020, Ayres_2021}. Given their widely differing scattering rates, Fermi surface (FS) geometries and magnetic field orientation dependencies, it seems unlikely that mechanisms involving quantum MR \cite{Abrikosov98}, sharp Fermi surface corners \cite{pippard} or Zeeman splitting \cite{Ando_2002, Hayes_NatPhys_2016, giraldogallo_science_2018} are responsible. 

Other candidate proposals fall essentially into two categories. The first, based on random resistor networks \cite{Parish03}, attributes LMR to the presence of disorder, through real space binary distributions \cite{boyd_prb_2019}, real space patches \cite{patel18} or doping inhomogeneity \cite{Singleton20}. Given the large variations in disorder levels and electronic inhomogeneity present in Bi2201, LSCO and Tl2201, however, it is difficult for such models to account for the similarity in form and magnitude of the MR seen in different families or the observed relation to intrinsic properties such as $T_c$.

The second category centers on models involving momentum space differentiation, specifically models where orbital motion is inhibited by some form of anisotropic impedance; the stronger the impedance, the higher the field range over which LMR is observed \cite{Hinlopen_2022}. While certain systems, e.g. LSCO and Nd-LSCO, appear to be consistent with this categorization due to a specific feature of their band structure (to be outlined below), such specificity once again conflicts with the universality of the MR observed across various cuprate families.

In highly-doped non-SC LSCO ($x \approx$ 0.30), a sizeable (10-fold) anisotropy in the low-$T$ quasiparticle mean-free-path $\ell_0(\phi)$ was inferred from earlier ARPES \cite{Yoshida_PRB_2006} and low-field Hall effect \cite{narduzzo_2008} measurements. Such marked anisotropy was attributed to the proximity of $\epsilon_F$ to the vHs at the zone boundary and the predominance of small-angle (elastic) scattering induced by out-of-plane substitutional disorder \cite{abrahams_2000}. As shown in Fig.~6 of the SI, the MR of non-SC LSCO also exhibits LMR at the highest fields. This, coupled with adherence to Kohler scaling ($\beta^{-1}(T) \propto \rho(0,T) \propto c + dT^2$ -- see Fig.~\ref{fig:lsco-scaling}b) suggests that the LMR observed in overdoped non-SC LSCO can be understood within a Boltzmann treatment incorporating strong low-$T$ anisotropy in $\ell_0(\phi)$.

In a recent angle-dependent MR study, Grissonnanche {\it et al.}~deduced an even stronger, 100-fold in-plane anisotropy in $\ell_0(\phi)$ in SC samples of Nd-LSCO (where $\epsilon_F$ is almost coincident with the vHs) \cite{grissonnanche_2021}. Incorporating the deduced form of $\ell_0(\phi)$ into their Boltzmann analysis, the authors were able to generate an in-plane MR response exhibiting LMR at high fields \cite{grissonnanche_2021}, as found experimentally \cite{ataei_2022}. It is not yet clear, however, whether the MR reported in Ref.~\cite{ataei_2022} obeys power-law scaling, as shown here for SC LSCO (Fig.~\ref{fig:lsco-scaling}a), or the Kohler-like scaling seen in non-SC LSCO. The precise form of the observed scaling is actually an important, though largely overlooked point, as we will argue below. It is also important to stress that strong residual anisotropy in $\ell_0(\phi)$ has not been reported in Bi2201 \cite{Kondo06} nor in Tl2201 \cite{Abdel_Jawad_2006, Peets_2007} despite both systems exhibiting an almost identical MR response to that seen in LSCO. Moreover, a recent Boltzmann-based model invoking $T$-dependent anisotropy in $\ell(\phi)$ was shown to be incapable of reproducing simultaneously the $H,T$-dependencies of the in-plane MR and Hall resistivity observed in Tl2201. Specifically, the level of anisotropy required to account for the magnitude of the MR led to a $T$- and $H$-dependent Hall resistivity that was wholly inconsistent with the experimental data. Finally, in both Bi2201 and Tl2201, the magnitude of the MR is seen to grow as each system is doped further away from the vHs (with reducing $p$) and thus any residual anisotropy from out-of-plane substitutional disorder must diminish. Taken together, these inconsistencies appear to rule out anisotropy in the elastic scattering channel as the dominant source of LMR, at least for $p^{\ast} < p < p_{sc}$. 

In fact, one can take this argument further and make a more general statement. Adopting a modified Boltzmann theoretical treatment, Hinlopen \textit{et al.}~recently showed that a high-field LMR of a similar magnitude to that observed can be a robust feature of any system in which cyclotron motion is impeded at discrete points or regions on the FS \cite{Hinlopen_2022}. Impeded cyclotron motion can be simulated by introducing an upper limit (\textit{bound}) on the time integral of the expression for the conductivity $\sigma_{xx}$ that makes the velocity-velocity correlation terminate at a specific boundary on the FS, i.e.

\begin{equation}
	\sigma_{xx} = \frac{e^2}{4\pi^3\hbar} \int_{FS} d^2k \int_0^{bound}\frac{v_i(0)}{v_F(0)} v_j(-t) e^{-t/\tau} dt
	\label{sigma_general_bounded}
\end{equation} 

\noindent where the velocity $v = \nabla_k\epsilon/\hbar$ and $\epsilon$ is the band energy. The model effectively covers multiple possible sources of impeded cyclotron motion such as partial gaps, nesting fluctuations or peaks in the density of states arising through hot spots \cite{Koshelev_PRB_2016}, turning points \cite{Koshelev_PRB_2013}, magnetic breakdown \cite{Naito_1982}, vHs \cite{grissonnanche_2021} or states on the Fermi surface whose charge dynamics have been rendered incoherent \cite{Putzke_NatPhys_2021}. Although the model is able to generate LMR, the absence of $\rho_0$ from the MR scaling could not be captured. This failure is particularly troublesome in Bi2201 where $\rho_0$ can be as high as 150 $\mu\Omega$cm, equivalent to an impurity scattering rate 1/$\tau_0$ of order several hundred Kelvin. Evidently, the observation of power-law scaling poses the most serious challenge to any semiclassical treatment of the MR in strange metals. The generality of the above model nevertheless compels us to consider impeded orbital motion as a possible origin of LMR and to seek potential sources of impedance across the phase diagram. 

As the pseudogap opens, states are removed from the Fermi level initially near the anti-nodal regions of the FS. This is believed to lead to the creation of Fermi arcs, that may, below a certain doping level $p_{CO} < p^{\ast}$ and in the presence of a large magnetic field, undergo FS reconstruction into a set of diamond-shaped Fermi pockets located near the zone diagonals \cite{Harrison_PRL_2011}. Both the arcs (discontinuities in the FS) and the pockets (sharp turning points) can create the necessary conditions for impeded cyclotron motion and thus LMR at the highest field strengths. Moreover, the reduction in density of states with further decrease in $p$ will lead to a shrinking of the arc lengths and/or the pocket dimensions, both of which could drive the observed increase in $\gamma$ with lowering $p$ inside the pseudogap regime. 

The origin of the impedance within the strange metal regime, where the underlying FS is believed to be intact, is arguably harder to identify. Given that $H/T$ scaling is tied inextricably to $T$-linear resistivity, the answer may lie in the emergence of non-quasiparticle states that have been rendered highly dissipative or incoherent via scattering off critical fluctuations \cite{wu_2021} or through some other mechanism \cite{zaanen_scipost_2019, Banerjee_2021}. According to Ref.~\cite{Hinlopen_2022}, $\gamma$ is determined by the size of the \lq impeded' region(s), in which case, the increase in both $\gamma$ and $a_1$ with underdoping could signify an increase in the width (in $k$-space) of these incoherent sectors. It is important to recall that the growth in $\gamma$ and $a_1$ occurs \textit{before} the pseudogap opens and tellingly, coincides with the $p$ to 1+$p$ crossover in the effective Hall number across the strange metal regime \cite{Putzke_NatPhys_2021}.  

Fluctuating charge order has been reported across the cuprate phase diagram \cite{arpaia_2021_charge,Uchida21}, including in the three families studied here \cite{Lin20, Peng_NatMat_18, Tam21}. Likewise, spin fluctuations are known to extend to the edge of the SC dome at $p = p_{sc}$ in LSCO \cite{Wakimoto_2007}. Were fluctuations of either type to be of sufficiently low-energy and sufficiently damped, scattering off such fluctuations could account for the persistence of $T$-linear resistivity down to low-$T$ \cite{caprara_2022}. To account simultaneously for LMR, however, they would have to give rise to residual (i.e. $T$ = 0) anisotropic scattering -- not foreseen in models based on short-range charge fluctuations \cite{caprara_2022} -- with an anisotropy that grows markedly with decreasing $p$. 

Finally, we turn to consider the field orientation dependence of the MR from the perspective of impeded orbital motion. In overdoped Bi2201 and Tl2201 \cite{Ayres_2021} as well as in underdoped Y123 \cite{harris_prl_1995}, the MR is relatively insensitive to field orientation \cite{Ayres_2021}, prompting suggestions that Zeeman effects were in fact playing a central role. Certainly, the quadrature expression given in Eq.~(\ref{eq:QuadT_2}) implies the summation of thermal and magnetic energy scales. In P-Ba122 \cite{Hayes_prl_2018}, FeSe$_{1-x}$S$_x$ \cite{licciardello_coexistence_2019}, optimally doped Y123 \cite{harris_prl_1995} and LSCO (see Fig.~7 of the SI and \cite{ataei_2022}), however, the quadrature MR is largely governed by the out-of-plane component of the field. In looking at possible explanations for this dichotomy, it is interesting to note that the resistive anisotropy in underdoped Y123 and in Bi2201 and Tl2201 is orders of magnitude larger than in P-Ba122, FeSe$_{1-x}$S$_x$, optimally doped Y123 and overdoped LSCO. Hence, while the origin of this distinction is not known, one might speculate that in the presence of an in-plane field, orbital motion in the more 2D cuprates is also impeded along $k_z$, giving rise to quadrature MR of a similar magnitude, while in more three-dimensional systems, such a tunneling barrier is absent. 

In summary, while certain aspects of the MR response within the strange metal regime, particularly those associated with the quadrature form of the MR, appear to be consistent with a modified Boltzmann approach based on impeded cyclotron motion, there are other aspects that cannot be explained within this picture. Thus, as it stands, a global theory for $H/T$ scaling remains elusive.

\begin{figure*}[!ht]
              \centering        
              \includegraphics{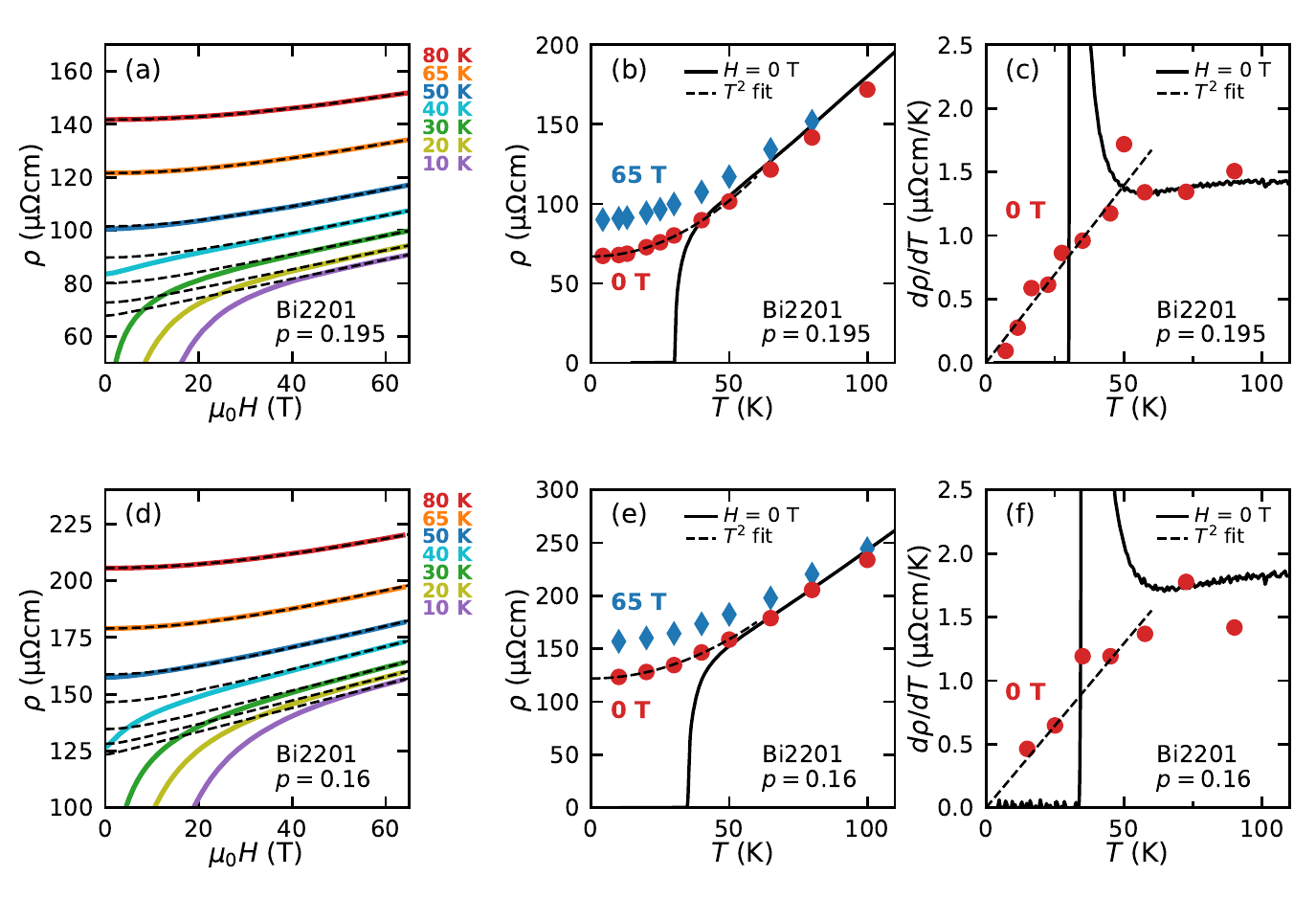}
              \caption{(a): Set of $\rho(H, T)$ curves for Bi2201 with $p$ = 0.195 (i.e. just below $p\ast$) at representative temperatures. Dashed lines are fits of the normal state data to Eq.~(\ref{eq:QuadT2_1}) that allow for the extrapolation of the high-field data to zero-field. (b): Measured zero-field resistivity (solid black line) of the same sample. High-field resistivity $\rho(T$, 65 T) determined from the end-points of the curves presented in (a) are denoted by blue diamonds whilst the extrapolated zero-field resistivity is denoted by red circles. The extrapolated $\rho(0,T)$ curve (red circles) obeys a $T^2$ dependence (dashed line fits). (c): Equivalently, a crude derivative d$\rho$/d$T$ can be obtained from the extrapolated $\rho(0, T)$ (red circles in panel (b)) yielding an extrapolated zero-field behavior below $T_c$ (red circles). The obtained points (red circles) match the measured d$\rho$/d$T$ (solid black line) above $T_c$ and confirm pure $T^2$ behavior below $T_c$ as indicated by a straight-line fit (dashed line). (d-f): The same data and analysis as in panels (a-c) is presented for optimally doped Bi2201 ($p$ = 0.16). Again, the low-$T$ dependence of $\rho(H=0,T)$ is found to be quadratic.}
              \label{fig:low-t-resistivity}
\end{figure*}

\subsection{Inside the pseudogap regime}

The most profound change in the MR response below $p^{\ast}$ is the crossover from $H/T$ to $H/T^2$ scaling. Intriguingly, a similar crossover from $H/T$ to $H/T^2$ scaling was also observed in P-Ba122. At its antiferromagnetic QCP, the MR in P-Ba122 exhibits $H/T$ quadrature scaling \cite{Hayes_NatPhys_2016}, while inside the magnetic phase, the MR scaling changes to $H/T^2$ coinciding with a crossover to $T^2$ resistivity \cite{Maksimovic_prx_2020}. The LMR in P-Ba122 was attributed to different origins within the two regimes: to scattering off hotspots on the FS at the QCP \cite{Koshelev_PRB_2016} and to FS reconstruction and the creation of sharp turning points inside the antiferromagnetic phase \cite{Koshelev_PRB_2013}.

While it is tempting to consider a similar scenario in the hole-doped cuprates, the continuation of $H/T^2$ scaling up to room temperature appears to rule out FS reconstruction (e.g. due to charge order) as its origin, since the latter is believed to occur only at low $T$. The step-wise change in the MR scaling is therefore more likely to be a consequence of the loss of states at the anti-nodes, its sharpness reminiscent of the abrupt change in the electronic specific heat \cite{Loram_2001} or the loss of coherent quasiparticle peaks in the (single-particle) spectral functions across $p^{\ast}$ \cite{chen_incoherent_2019}. Given the step-wise change in scaling, it is possible that the intermediate power-law exponents recorded within the crossover region reflect a (real-space) coexistence of gapped and non-gapped regions exhibiting $H/T^2$ and $H/T$ scaling, respectively, though more detailed studies around $p^\ast$ would be needed to confirm whether this is indeed the case.

Given that the scaling is observed over an extended range of temperatures and fields, we can use its form and the exponent $m$ to extrapolate the MR to $H\to 0$ for $T<T_c$ and thus extract $\rho(0,T)$ in the absence of superconductivity. In panels (a) and (d) of Figure \ref{fig:low-t-resistivity}, the MR data of two Bi2201 crystals with $p$ = 0.195 and 0.16, respectively, are fitted to Eq.~(\ref{eq:QuadT2_1}) with $m$ fixed to the value found through the scaling plots described above and $\alpha_m$ and $\gamma$ determined by fits to the scaled derivatives. For the entire data set that can be ascribed to the normal-state, the fits are excellent. In panels (b) and (e) of Fig.~\ref{fig:low-t-resistivity}, the corresponding high-field (65 T) and extrapolated ($H \to 0$) values of the resistivity are plotted. At low $T$, $\rho(0,T)$ clearly deviates from its higher temperature $T$-linear behavior. Finally, in panels (c) and (f), the derivatives d$\rho(0, T)$/d$T$ are plotted. Around $T$ = 60 K, the zero-field d$\rho/$d$T$ diverges due to the onset of SC fluctuations. In red, the derivatives of the extrapolated resistivity reveal that d$\rho/$d$T \to 0$ in a linear fashion, indicative of a crossover to a $T^2$ dependence of $\rho(0,T)$.

The analysis presented in Fig.~\ref{fig:low-t-resistivity} reveals that in optimally doped cuprates or in samples with $p \lesssim p^{\ast}$, $\rho(0,T)$ is not $T$-linear, as is widely believed. While low-$T$ $T^2$ resistivity has been reported at lower dopings \cite{Rullier_Albenque08, chan_prl_2014, proust_PNAS_2016}, we are not aware of any such report so close to $p^{\ast}$ or at optimal doping. A note of caution is required here, however. Firstly, the obtained quadratic behavior in Bi2201 contrasts with the predominantly $T$-linear resistivity observed in LSCO just below $p^{\ast}$ (also extrapolated from high-field MR measurements) \cite{giraldogallo_science_2018}. Moreover, small resistive upturns are often seen in Bi2201 crystals at low $T$ \cite{Ono_2000, Lizaire_2020, Berben_PD_2022} that may mimic a crossover to $T^2$ resistivity. This seems unlikely however given the fact that in both LSCO and Bi2201 the MR exhibits pure $H/T^2$ scaling below $p^{\ast}$ indicating a significant $T^2$ component in $\rho(0, T)$ that is itself subject to impeded cyclotron motion in an applied magnetic field. A conclusion to draw from this analysis is that the only region of the phase diagram where a $T$-linear component to the resistivity survives to the lowest temperatures is the strange metal regime $p^{\ast} < p < p_{sc}$. This is also the only regime where $H/T$ scaling of the MR is manifest. 

The precise location of $p^{\ast}$ in Bi2201 has been the subject of much debate \cite{Ando_2004_curvature, Zheng_2005, kondo_2011_disentangling, Lizaire_2020, Berben_PD_2022}. NMR studies have suggested that $p^{\ast}$ lies close to the edge of the SC dome \cite{Zheng_2005}, though the distinction between $T^*$ and $T_{\mathrm{F}}$ has not yet been fully resolved. A more recent study involving transport and ARPES measurements found that the pseudogap opens around $p=0.2$ \cite{Berben_PD_2022}. The MR results presented above shed new light on this issue by uncovering an abrupt change in the scaling of the MR from $H/T$ to $H/T^2$ at $p \sim 0.2$. Given that (a) $H/T^2$ scaling is seen in different underdoped cuprates (such as LSCO, Hg1201 and Bi2201) and (b) a similar crossover in the MR occurs in LSCO across $p^{\ast}$ ($\sim$ 0.2), the crossover observed in Bi2201 at $p \sim$ 0.2 appears to confirm that manifestations of the pseudogap in the charge transport end here.

Another key finding of this study is that for samples just below $p^{\ast}$, no distinction in the scaling of the MR can be made for temperatures below and above $T^*$. While the zero-field resistivity departs from a pure linear form below $T^*$ (for $p$ = 0.195, for example, $T^* \sim$ 150-200 K), a single power-law can describe the MR over the whole temperature range from 300 K to 4 K. These findings tend to support the notion that the pseudogap is controlled by an energy scale rather than a phase transition \cite{chen_incoherent_2019,Tallon_PRB_2020}.
 
Finally, despite the observation of quantum oscillations \cite{doiron_leyraud_07}, the verification of the Wiedemann-Franz law \cite{Grissonnanche_PRB_2016,Michon_PRX_2018}, the recovery of a $T^2$ resistivity \cite{proust_PNAS_2016,chan_prl_2014} and a $\omega^2$ scattering rate \cite{mirzaei_spectroscopic_2013} inside the pseudogap regime, the appearance of Fermi arcs challenges claims of a conventional FL ground state below $p^\ast$. The extended range of $T^2$ resistivity is also anomalous. An in-plane resistivity of the form $c + dT^2$ is limited, for example, to temperatures $\leq$ 160 K \cite{chan_prl_2014}, while the pure $H/T^2$ scaling of the MR presented here is found to persist up to room temperature. The manifestation of pure power-law scaling is also at odds with both conventional and modified Boltzmann transport theory, suggesting that the presence of a robust $T^2$ resistivity may require an explanation that is beyond the realm of standard FL theory \cite{Reber2019}. 

\section{Conclusion}

In the present study, we have carried out a universal analysis of the temperature, field and doping dependence of the in-plane MR of hole-doped cuprates to reveal unique scaling behavior within each distinct regime of the phase diagram: $H/T^2$ scaling in the pseudogap regime, $H/T$ scaling in the strange metal regime and Kohler scaling in the non-SC FL regime. In addition, we have provided definitive answers to the four outstanding questions highlighted in the Introduction:

{\bf A:} $H$-linear MR is observed at the highest fields in all SC samples studied, with a magnitude that falls monotonically with increasing $p$, extrapolating to zero around $p_{sc}$ in precisely the same manner as $a_1$, the $T$-linear coefficient of $\rho(0,T)$. The origin of the variation of $\gamma$ above and below $p^{\ast}$, however, is argued to be distinct. Nonetheless, this correlation shows that the magnetotransport connects normal-state room temperature physics with $T_c$, even when the two differ by two orders of magnitude. Moreover, the ubiquity of the $H/T$ scaling throughout the entire strange metal regime adds further weight to its interpretation as a quantum critical phase rather than a response driven by proximity to a singular critical point.

{\bf B:} Across $p^\ast$, the form of scaling changes abruptly from $H/T$ to $H/T^2$ scaling in the pseudogap regime signifying a loss of the $T$-linear component in both $\rho(0,T)$ and $\Delta \rho(H,T)$ below $p^\ast$. Nevetheless, the persistence of power-law scaling inside the pseudogap regime suggests that the remnant states near the Fermi level still possess non-FL character. As in the strange metal regime, $\rho_0$ and, by default, $1/\tau_0$ are both absent from the scaling, even when $1/\tau_0$ is of order 100-200 K. Data from four different cuprate families are found to exhibit the same power-law scaling, demonstrating remarkable universality of the phenomenon. Finally, the marked change in MR scaling mirrors that seen in the single-particle response \cite{Loram_2001, chen_incoherent_2019}; the lack of signatures of $T^\ast$ in the MR supporting interpretations of the pseudogap as a doping-dependent energy scale that vanishes at $p^{\ast}$, rather than a phase transition to some form of hidden order.

{\bf C:} In overdoped non-SC LSCO, beyond $p_{sc}$, the low-field MR follows more conventional Kohler scaling with $\beta^{-1}(T) = c + dT^2$. The recovery of a finite residual component in $\beta^{-1}(T)$ across $p_{sc}$ appears to be similarly abrupt. Despite the recovery of Kohler scaling (at low fields), the MR still shows a tendency towards $H$-linearity at the highest fields. We have argued here that the observation of LMR in LSCO reflects the presence of a highly anisotropic elastic mean-free-path $\ell_0(\phi)$ \cite{narduzzo_2008}. In highly doped, non-SC LSCO, the LMR is essentially \lq accidental' and we expect measurements to higher fields will show a tendency towards saturation as the anisotropy in $\ell_0(\phi)$ is eventually washed out. At lower doping levels, corresponding to where $\epsilon_F$ crosses the vHs, the anisotropy in $\ell_0(\phi)$ may become so large \cite{grissonnanche_2021} that the LMR will persist up to the highest accessible field strengths. The observation of residual-free scaling in SC samples, however, is striking and cannot be captured by any modified Boltzmann theory and clearly requires further experimental and theoretical study to be resolved.

{\bf D:} While modified Kohler scaling (Eq.~(\ref{eq:MKR})) is observed in both the pseudogap and strange metal regimes, the ubiquity of power-law scaling (Eq.~(\ref{eq:QuadT2_2})) and its validity up to the highest fields clearly identifies the latter as the intrinsic field-induced response of SC hole-doped cuprates. As a result, modified Kohler scaling is relegated to an empirical finding generated by dividing $\Delta\rho(H)$ by $\rho(0)$. How the Hall and MR responses can be reconciled within the LMR regime, however, remains to be resolved.

While the generation of LMR at higher field strengths is compatible with a scenario based on impeded cyclotron motion, the observed power-law scaling is not. Nevertheless, one can speculate as to the origin of any such impedance within the various doping regimes. As mentioned above, in highly-doped non-SC LSCO, LMR may arise from a density of states effect amplified by the presence of out-of-plane substitutional disorder. In the pseudogap regime, the termination of the Fermi arc provides a natural \lq boundary' to cyclotron motion, one that is unlikely to be penetrated at higher field strengths. Thus, we predict that the LMR in pseudogapped cuprates will not deviate up to the highest fields available. The origin of LMR within the strange metal regime is arguably the most difficult to discern. Nevertheless, the close correlation between $\gamma$ and $a_1$ suggests that its origin is tied to the emergence of non-FL transport and the accompanying loss of coherent quasiparticle states. Within this picture, the growth in $\gamma$ across the strange metal regime implies an expansion of this incoherent sector with decreasing $p$, as inferred from our earlier Hall study \cite{Putzke_NatPhys_2021}. 

When $p$ falls below $p^{\ast}$ (in Bi2201), $\gamma$ continues to increase in a linear fashion, reflecting the reduction in Fermi arc length as the pseudogap grows. In Bi2201, the extrapolated low-$T$ resistivity becomes purely quadratic as $T\to0$ just as the quadratic MR develops $H/T^2$ scaling, indicating a suppression of the $T$-linear component in $\rho(0,T)$ at low-$T$. Knowledge of the anisotropy of the pseudogap then leads us to infer that the states removed from the Fermi level below $p^{\ast}$ were in fact those exhibiting signatures of incoherent transport inside the strange metal regime.  Finally, the sharpness of the crossover in power-law scaling from $H/T$ to $H/T^2$ is reminiscent of the abrupt changes seen in earlier ARPES \cite{chen_incoherent_2019} and specific heat \cite{Loram_2001} studies.

In closing, we return to the central conundrum of our study -- the fact that the MR exhibits pure power-law scaling with no residual component (as would have been expected from conventional or modified Kohler scaling). This robust experimental result suggests that the magnetotransport properties of superconducting cuprates require an altogether new theoretical framework, one that is incompatible with conventional Boltzmann transport theory and entirely decoupled from impurity scattering. The direct correlation between $H$-linear MR, $T$-linear resistivity and $T_c$ suggests that developing such a framework will surely pave the way for a more unified description, not only of the anomalous normal state of hole-doped cuprates, but also of high-$T_c$ superconductivity itself.
\\

\section{Methods}

High quality crystals of Bi2201 and LSCO were grown in floating zone furnaces located at different sites. To cover a large part of the phase diagram, some of the Bi2201 crystals were doped with La and Pb, resulting in the chemical formula: Bi$_{2+z-y}$Pb$_y$Sr$_{2-x-z}$La$_x$CuO$_{6+\delta}$. The Tl2201 crystal studied here was synthesized using a self-flux technique \cite{Tyler1997}. The hole doping for Bi2201 was estimated from the measured $T_c$ (defined as the temperature at which the zero-field resistivity vanishes below the noise floor of our experiments) using the Presland relation \cite{presland_general_1991}: $1-T_c$/$T^{\rm max}_c$ = 82.6($p$ -- 0.16)$^2$ with $T^{\rm max}_c$ = 35 K. This relation was recently demonstrated to hold well in Bi2201 \cite{Berben_PD_2022}. For LSCO, the doping was estimated from the measured $T_c$ using the same parabolic relation with $T^{\rm max}_c$ = 38 K and found to match closely the Sr content of each crystal. The crystallographic axes of LSCO were oriented with a Laue camera. Typical sample dimensions were 1000 $\times$ 250 $\times$ 10 $\mu$m$^3$ for Bi2201 and Tl2201 and 1500 $\times$ 250 $\times$ 50 $\mu$m$^3$ for LSCO. The $\rho(0,T)$ curves of all the MR samples are shown in Figure 1 of the SI. 

The MR was measured in DC and pulsed magnets up to 35 T and 70 T, respectively with the current $I$ applied in-plane and \textbf{H} $\parallel c$. For the pulsed field measurements, samples and wires were fully covered in GE varnish and/or vacuum grease to reduce vibration. To increase the measurement signal, each Bi2201 crystal was mechanically thinned to a thickness of 2 - 10 $\mu$m, resulting in sample resistances of $\sim$ 1 $\Omega$, i.e.~comparable to the resistance of the current contacts. At each measurement temperature, the MR curves were recorded for both polarities of the magnetic field. For Tl2201, a single crystal with an ambient pressure $T_c$ = 35 K was selected and prepared for transport measurements under the application of hydrostatic pressure using a piston cylinder cell. The sample was oriented on a feed-through such that the magnetic field \textbf{H} $\parallel c$. Daphne 7373 oil was used as a pressure transmitting medium as it is known to remain hydrostatic at room temperature (the temperature at which pressure was applied) up to 2.2 GPa \cite{Yokogawa2007}, beyond the pressures applied in this work.

%

%
%
%

\newpage
\setcounter{section}{0}
\setcounter{figure}{0}
{\large Supplementary Information}

\section{Description and characterization of all samples under study}

\begin{table*}[h!]
\centering
\begin{tabular}{p{0.8cm} p{0.8cm} p{0.8cm} p{0.8cm} p{3.8cm} p{1.2cm} p{0.9cm} }
\textbf{Bi2201} & \\
Bi & Pb & Sr & La & Annealing & Doping & $T_c$ \\
\hline
1.72 & 0.38 & 1.85 & 0 & 400C, 96h in 2.5 atm. O$_2$ & 0.270 & $<$1 K \\
1.72 & 0.38 & 1.85 & 0 & 750C, 24h in Air & 0.260 & 7 K \\
1.72 & 0.38 & 1.85 & 0 & 550C, 72h in N$_2$ & 0.245 & 13 K \\
1.72 & 0.38 & 1.85 & 0 & 550C, 72h in N$_2$ & 0.240 & 17 K \\
1.72 & 0.38 & 1.45 & 0.4 & As grown & 0.22 & 24 K \\
1.35 & 0.85 & 1.47 & 0.38 & 600C, 24h in Air & 0.205 & 28 K \\
1.2 & 0.9 & 1.3 & 0.55 & As grown & 0.195 & 30 K \\
1.35 & 0.85 & 1.47 & 0.38 & 650C, 72h in N$_2$ & 0.16 & 35 K \\
1.2 & 0.9 & 1.3 & 0.55 & 650C, 72h in N$_2$ & 0.13 & 31 K \\
\hline	
\end{tabular}
\caption{Details of the Bi2201 samples measured as part of this study. The $T_c$ values are defined as the temperature below which the zero-field resistivity falls below the noise floor. The doping levels are then determined from the $T_c$ values using the Presland formula \cite{presland_general_1991} with $T_c^{\rm max}$ = 35 K. The Bi2201 sample labelled with a $T_c <$ 1 K was measured down to 1.4 K, and although it shows an onset of superconductivity, it does not become fully superconducting. We have therefore specified its doping level as $p = 0.27$ in line with the parabolic Presland formula.}
\end{table*}

\begin{table}[h!]
\centering
\begin{tabular}{p{2.8cm} p{2.8cm} p{2.8cm} }
\textbf{LSCO} & & \\
Sr content & $T_c$ & Inferred Doping\\
\hline
0.20 & 30 K & 0.205\\
0.23 & 22.4 K & 0.230\\
0.30 & no SC & / \\
0.33 & no SC & / \\
\hline	
\end{tabular}
\caption{Details of the LSCO samples studied. The $T_c$ values are defined as the temperature below which the zero-field resistivity falls below the noise floor. The listed Sr content values are either the nominal ones or, for $x$ = 0.33, have been measured with electron-probe micro-analysis \cite{nakamae_prb_2003}. The samples with $x$ = 0.30 and 0.33 do not show any sign of superconductivity down to $T$ = 0.45 K.}
\end{table}

\begin{table}[h!]
\centering
\begin{tabular}{p{2.8cm} p{2.8cm} }
\textbf{Tl2201} &  \\
$T_c$ & Inferred Doping\\
\hline
35 K & 0.26\\

\hline	
\end{tabular}
\caption{Tl2201 sample studied as part of this work. The $T_c$ value is defined as the temperature below which the zero-field resistivity falls below the noise floor at ambient pressure. The same is true also for all data sets recorded under applied hydrostatic pressure. The doping level is inferred from $T_c$ as described in Ref.~\cite{Putzke_NatPhys_2021}.}
\end{table}

\begin{figure}[h!]
              \centering        
              \includegraphics[width=0.8\textwidth]{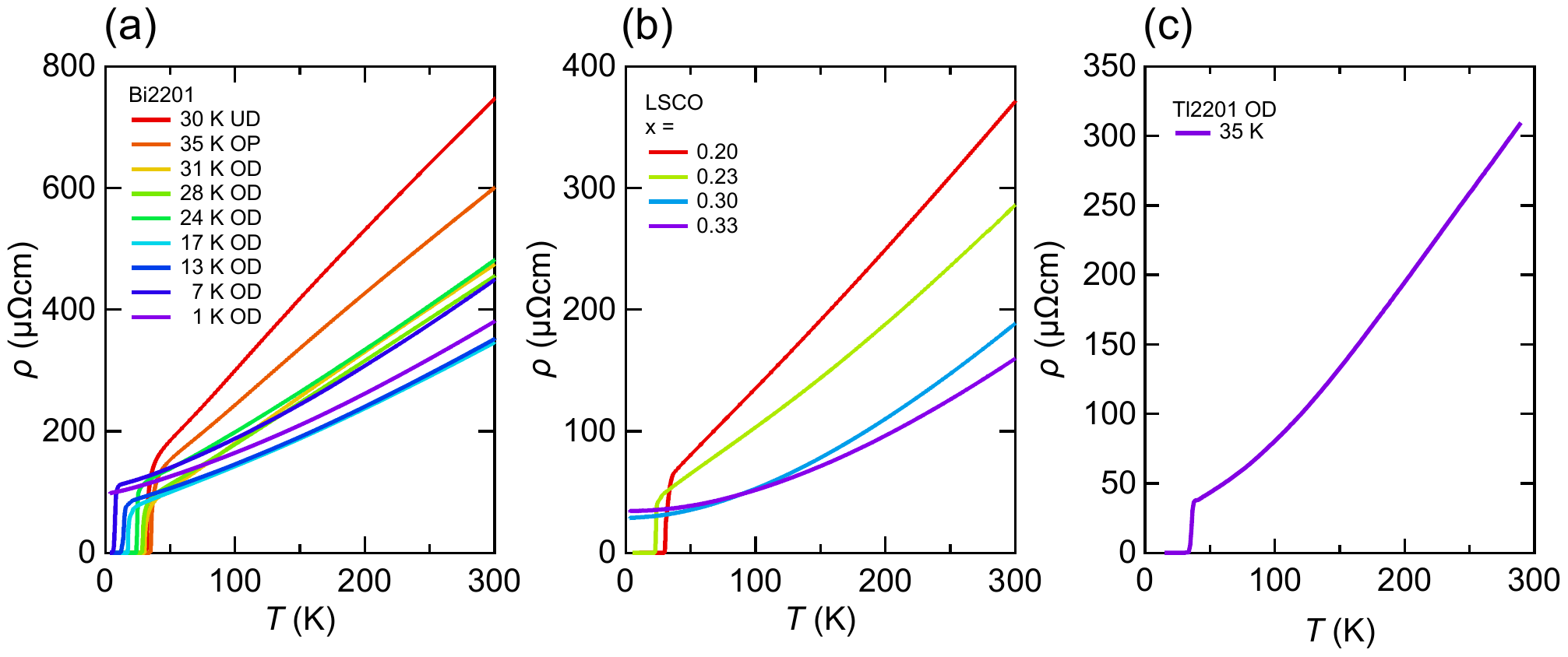}
              \caption{Zero-field resistivity curves for the (a) Bi2201, (b) LSCO and (c) Tl2201 samples whose magnetoresistance was investigated as part of this study. The error in the absolute values of $\rho(0,T)$ (due to geometrical uncertainties) is estimated to be $\sim$ 15\%.}
              \label{fig:zero_field_resistivity}
\end{figure}

\newpage

\section{Magnetoresistance of overdoped Tl2201 under pressure}

\begin{figure}[h!]
              \centering        
              \includegraphics{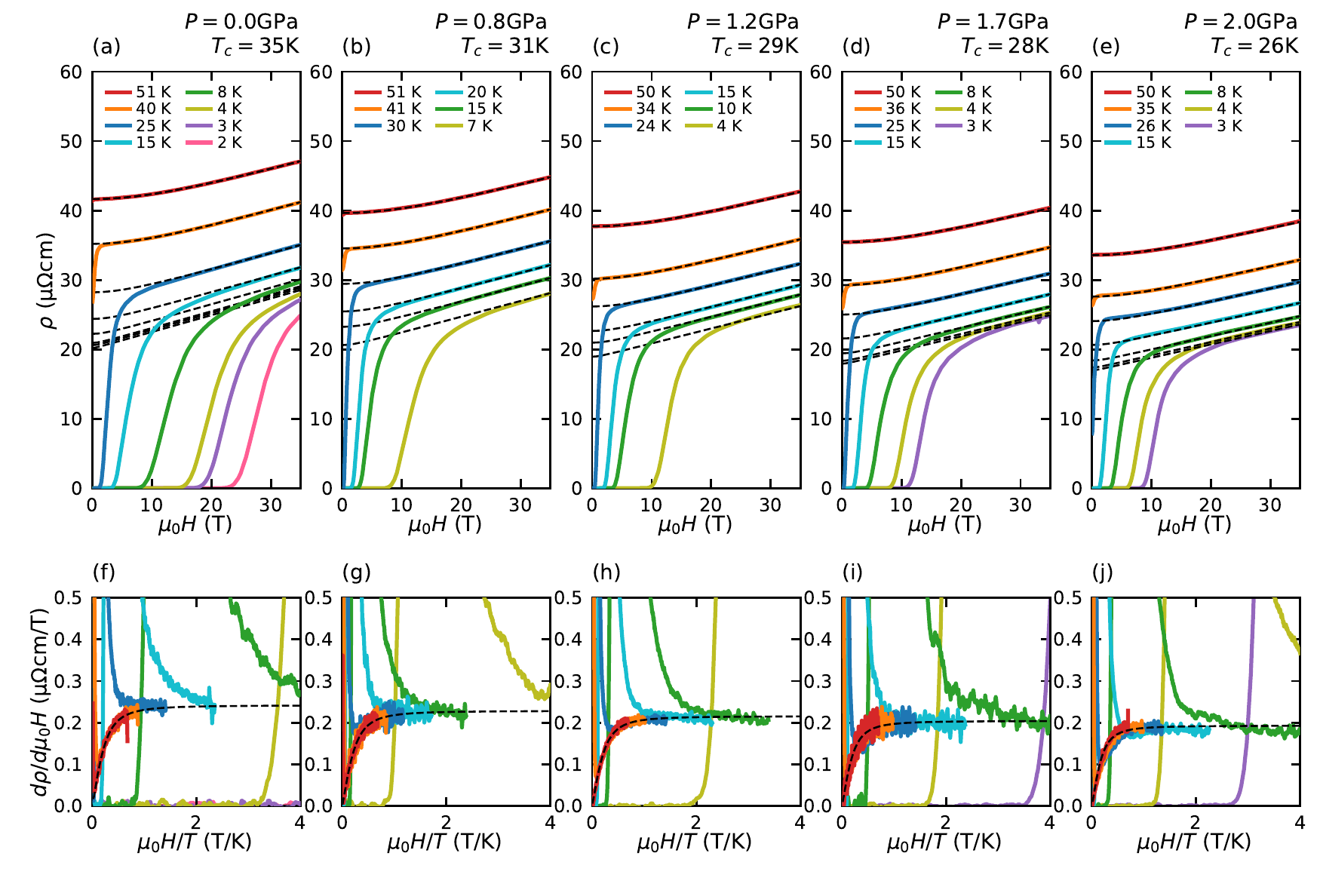}
              \caption{(a)-(e) Evolution with pressure of $\rho(H, T)$ of the Tl2201 single crystal (ambient pressure $T_c$ = 35K). Dashed lines are simultaneous fits to the MR to Eq.~(3) of the main report. (f)-(i) Derivatives d$\rho$/d($\mu_0H$) of the MR presented in panels (a)-(e). The derivatives presented in panels (f)-(i) are plotted against $H/T$. The dashed lines are derived from a single fit to the full set of MR derivatives (in the normal state) to Eq.~(3) of the main article. $\gamma$ -- the limiting low-$T$, high-$H$ slope of the $H$-linear MR -- can be read off from the plateau region of the individual dashed line fits.}
              \label{fig:pressure-data}
\end{figure}
 
\newpage

\section{Quadrature scaling of the MR in overdoped LSCO ($x$ = 0.23)}
 
\begin{figure}[h!]
              \centering        
              \includegraphics[width=1\textwidth]{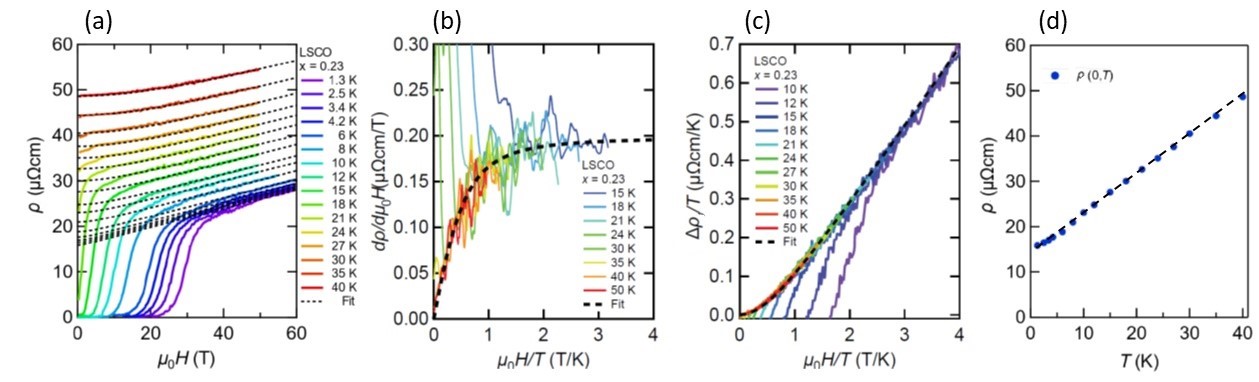}
              \caption{(a) Reproduction of the set of MR curves for LSCO $x$ = 0.23 reported in Ref.~\cite{cooper_science_2009}. Dashed lines are fits to the quadrature MR expression given in Eq.~(3) of the main report. (b) Corresponding plot of d$\rho$/d$\mu_0H$ vs.~$H/T$ indicating that the form of the MR does indeed follow $H/T$ scaling. Dashed line is the fit to the derivative of Eq.~(4) of the main article. (c) Corresponding plot of $\Delta\rho/T$ vs.~$H/T$. Again dashed line is the fit to Eq.~(3) of the main article. (d) Resultant $\rho(0,T)$ obtained by extrapolating the quadrature fits to the MR curves in panel (a) to $H$ = 0. Here, the dashed line is simply a guide to the eye.}
              \label{fig:Cooper_data}
\end{figure}

\newpage

\section{Simulation of quadrature scaling}

\begin{figure}[h]
              \centering        
              \includegraphics[width=0.7\textwidth]{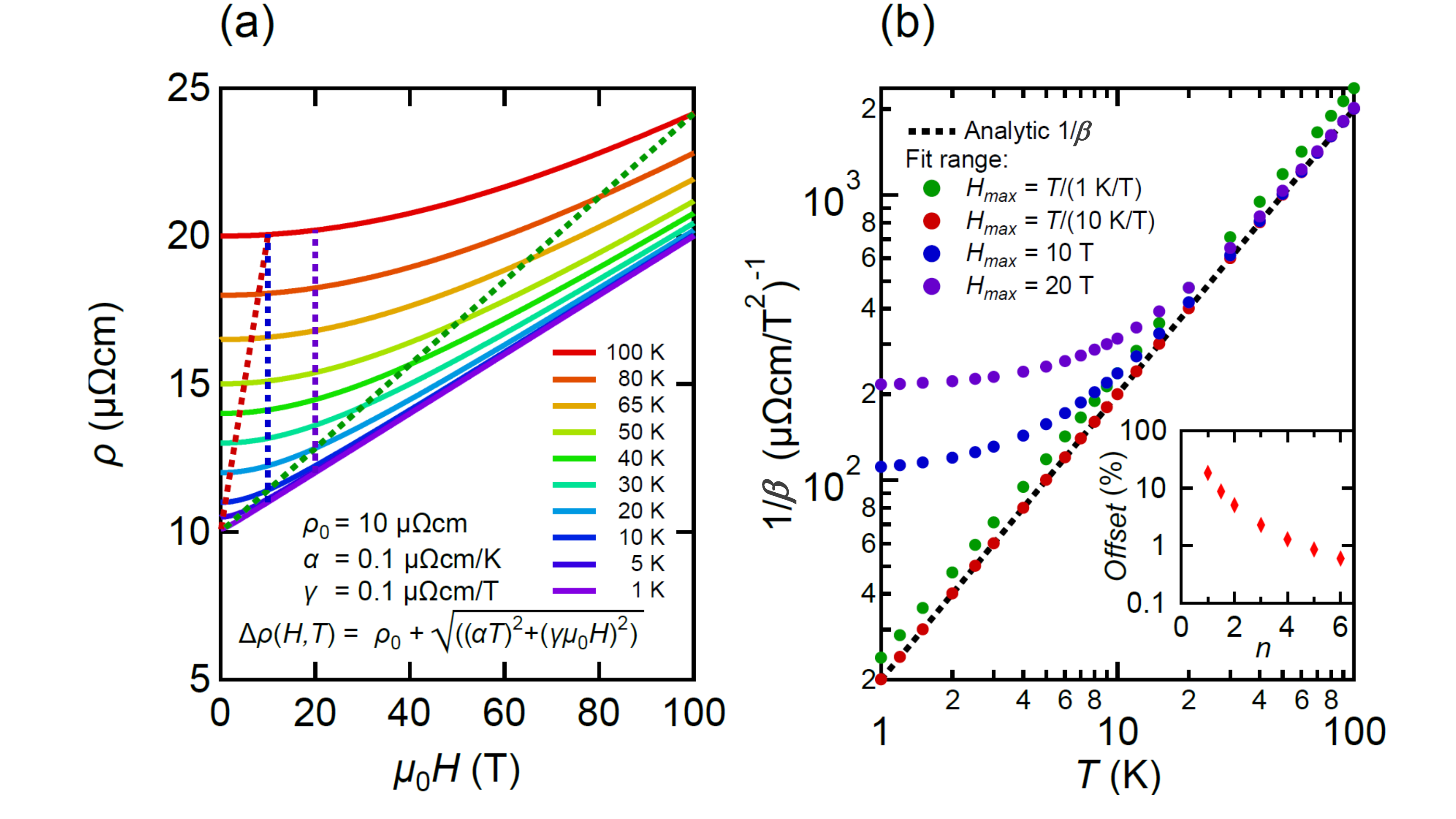}
              \caption{(a) Simulation of a set of MR curves obeying strict quadrature ($H/T$) scaling. Solid lines are the simulated dataset. The dashed lines indicate the fitting range used to fit each curve with a quadratic function. (b) $T$-dependence of the estimated coefficient $\beta$ of the low-field $H^2$ MR. Dashed black line indicates the analytical $H \to 0$ value of $\beta$. The circles indicate $\beta$ as found from the fitting procedure. The inset shows the offset of the analytical $\beta$ value in percent as described in the text.}
              \label{fig:MR_simulation}
\end{figure}

Since the hyperbolic form of the MR (Eq.~(4) in the main article) is never purely quadratic except in the $H \rightarrow$ 0 limit, fitting the data with a quadratic function is merely an approximation. In order to check whether our fitting procedure can capture a similar quadratic component to the analytical zero-field limit, we simulate in Figure 4 a dataset that obeys pure quadrature scaling, where the following parameters have been used: $\rho_0$ = 10 $\mu\Omega$cm, $\alpha$ = 0.1 $\mu\Omega$cm/K and $\gamma$ = 0.1 $\mu\Omega$cm/T. From Eq.~(5) of the main article, an analytical quadratic component $B$ = 0.05 $\mu\Omega$cm/K$^2$·$T$is found, as indicated by the black dashed line in Fig.~4(b).

In our approach, the simulated curves are first fitted with a quadratic function in a field range of 0 - 20 T and 0 - 10 T. While at high temperatures this gives a similar quadratic component to the analytical value, both start to deviate at low $T$. As expected, with a decrease of the fitting range, the deviation for the analytical value decreases at low-$T$. Given the $H/T$ scaling form of the quadrature MR and using the fact that $\gamma/\alpha$ represents the turnover field from quadratic to linear, we now use a fitting range that is determined in terms of $H/T$. With a range from 0 - ($\alpha T/\gamma$) Tesla (for the simulated data $\gamma/\alpha$ = 1 K/T) a constant offset of $\sim$ 18\% is found. While this is non-negligible, the key point is that the temperature dependence of the extracted $\beta$ coefficient remains $\sim 1/T$, and in the main article, it is the $T$-dependence that we are most interested in. Upon decreasing the $H/T$ range the offset quickly drops, reaching $\sim$ 1\% with a range of $\alpha T/ 5\gamma$, as shown in the inset of Fig.~4(b).

\newpage

\section{Power-law scaling of the MR in other cuprates}

\begin{figure}[h!]
              \centering        
              \includegraphics[width=0.6\textwidth]{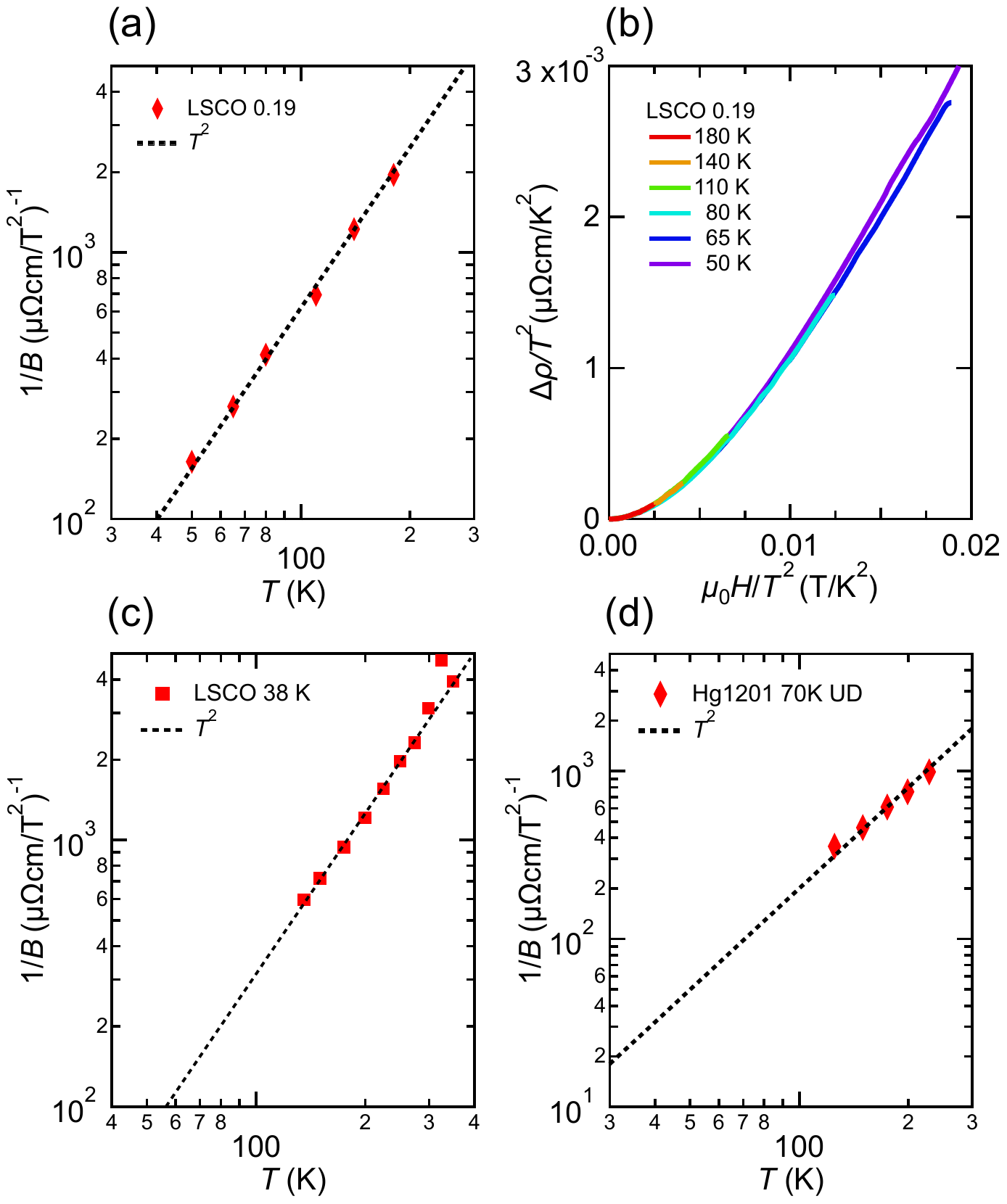}
          \caption{(a) 1/$B$ vs.~$T$ derived from the LSCO sample with $p$ = 0.19 reported by Giraldo-Gallo {\it et al.} \cite{giraldogallo_science_2018}. (b) Corresponding plot of $\Delta\rho/T^2$ vs.~$\mu_0H/T^2$ for the $p$ = 0.19 LSCO sample from Ref.~\cite{giraldogallo_science_2018}. (c) $1/B(T)$ for optimally doped LSCO from Harris {\it et al.} \cite{harris_prl_1995}. Here the zero-field resistivity curve needed to extract the $B$-term was taken from Kimura {\it et al.}. \cite{Kimura96}. (d) $1/B(T)$ for underdoped Hg1201 ($p$ = 0.14) from Chan {\it et al.} \cite{chan_prl_2014}. The dotted lines in panels (a), (c) and (d) are all guides to the eye.}
              \label{fig:Other_data}
\end{figure}

\newpage

\section{Resistivity and MR in highly overdoped non-SC LSCO}

\begin{figure}[h!]
              \centering        
              \includegraphics[width=0.7\textwidth]{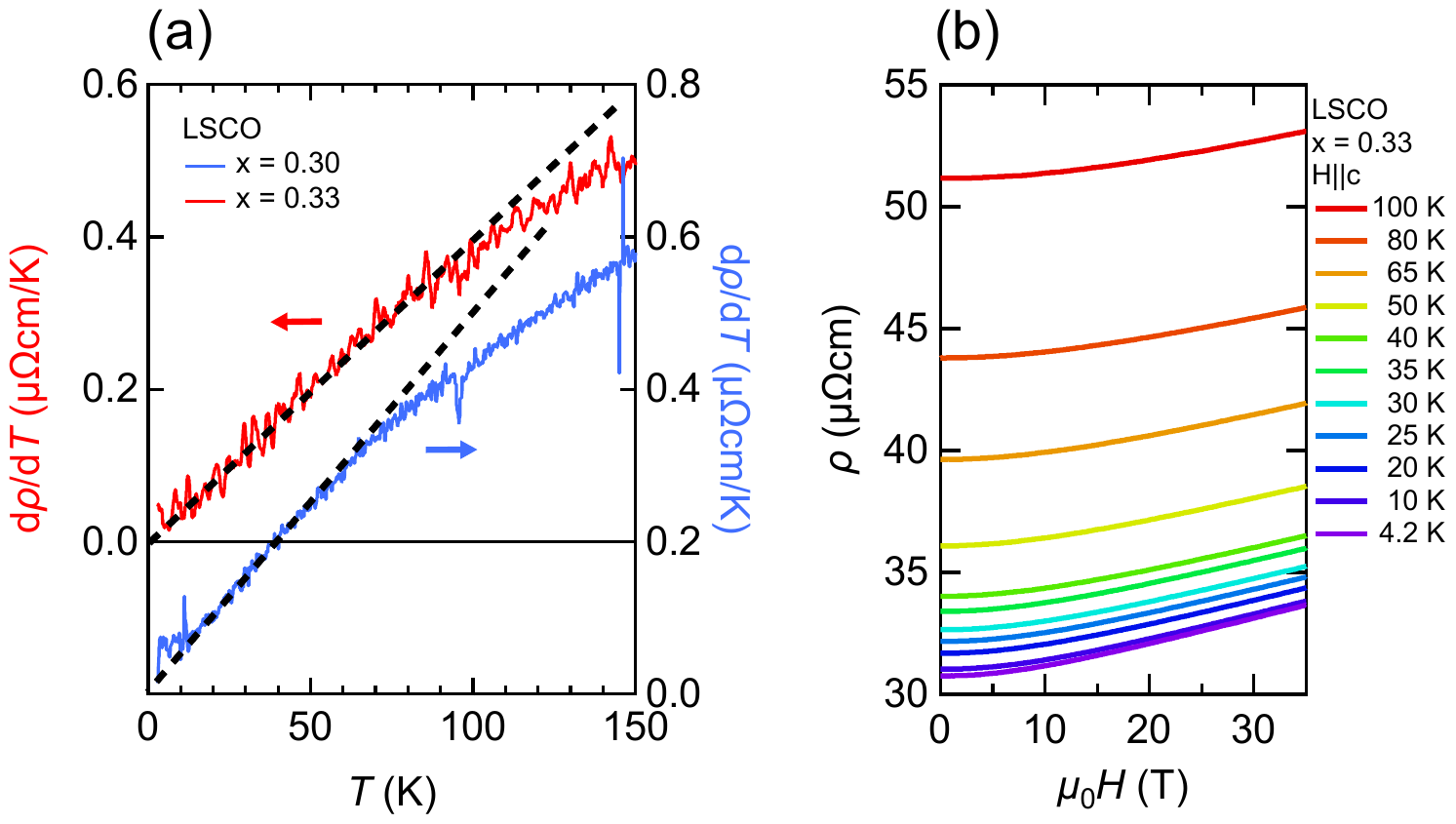}
         \caption{(a) Derivatives d$\rho$/d$T$ of the zero-field resistivity curves for the two highly overdoped non-superconducting LSCO samples shown in Fig.~\ref{fig:zero_field_resistivity} of the SI. The dashed lines are guides to the eye, highlighting the fact that $\alpha_1(0)$ - the low-$T$ $T$-linear resistivity coefficients (the intercepts on these derivative plots) is negligible and thus $\rho(T)$ shows purely quadratic behavior at low-$T$ consistent with correlated Fermi-liquid behavior. (b) Evolution  of $\rho(H, T)$ for $p$ = 0.33 up to 35 T between 100 K and 4.2 K. At low-$T$, a crossover to $H$-linear MR can be seen at the highest fields.}
              \label{fig:OD_LSCO}
\end{figure}

\newpage
\section{Comparison of Kohler and $H/T^m$ scaling in LSCO}

\begin{figure}[h!]
              \centering        
              \includegraphics[width=0.90\textwidth]{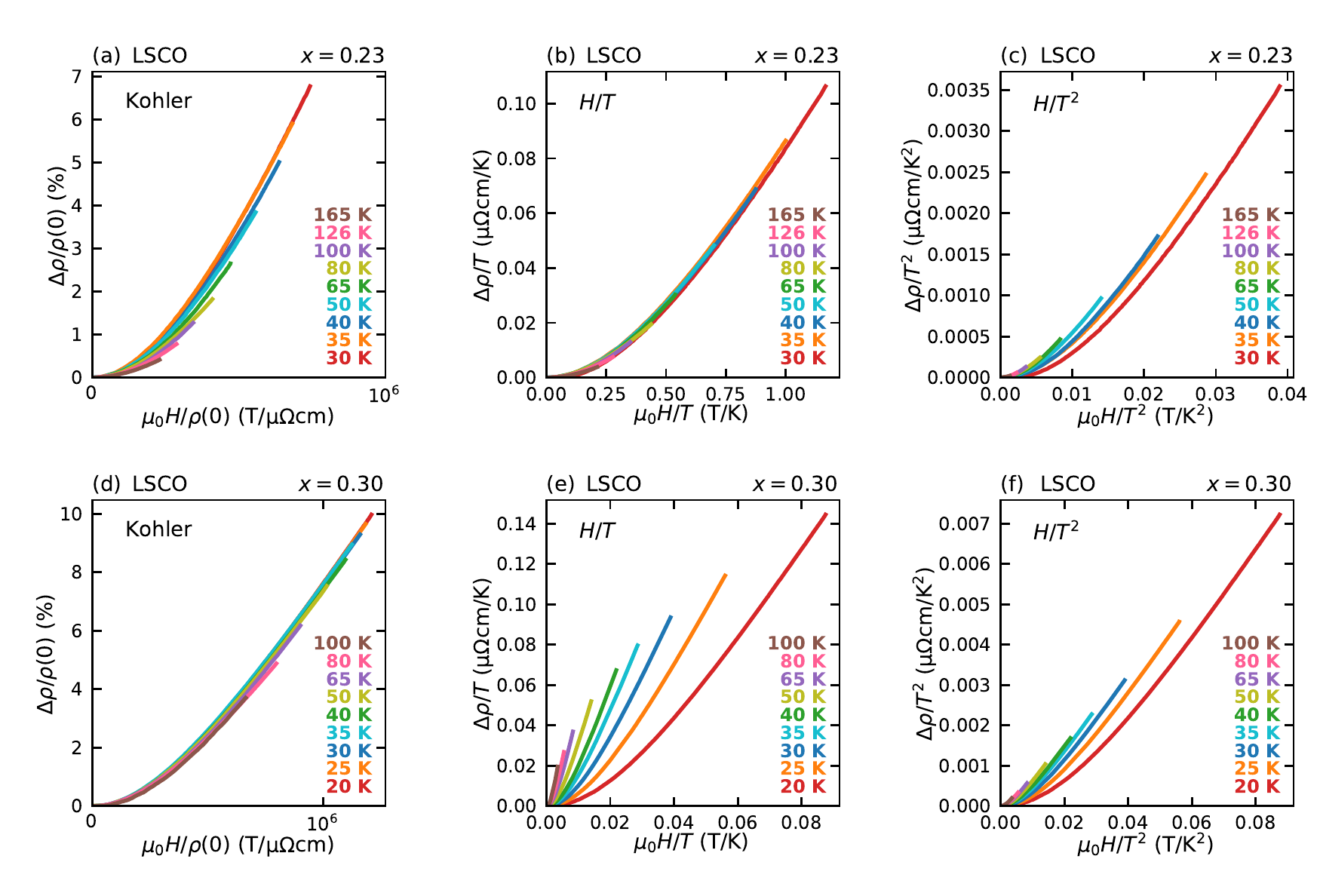}
         \caption{(a, b, c) A direct comparison of (a) Kohler, (b) $H/T$ and (c) $H/T^2$ scaling for LSCO with $x = 0.23$ within the strange metallic regime. In this doping regime, $H/T$ scaling is observed whilst both Kohler and $H/T^2$ scaling can be seen to fail. (d, e, f) The same scaling plots are also presented for non-superconducting LSCO with $x=0.30$, a doping regime in which Kohler scaling is most closely adhered to.}
              \label{fig:LSCO_kohler_vs_powerlaw}
\end{figure}

\newpage

\section{Anisotropic MR in overdoped LSCO ($x$ = 0.23)}

\begin{figure}[h!]
              \centering        
              \includegraphics[width=0.65\textwidth]{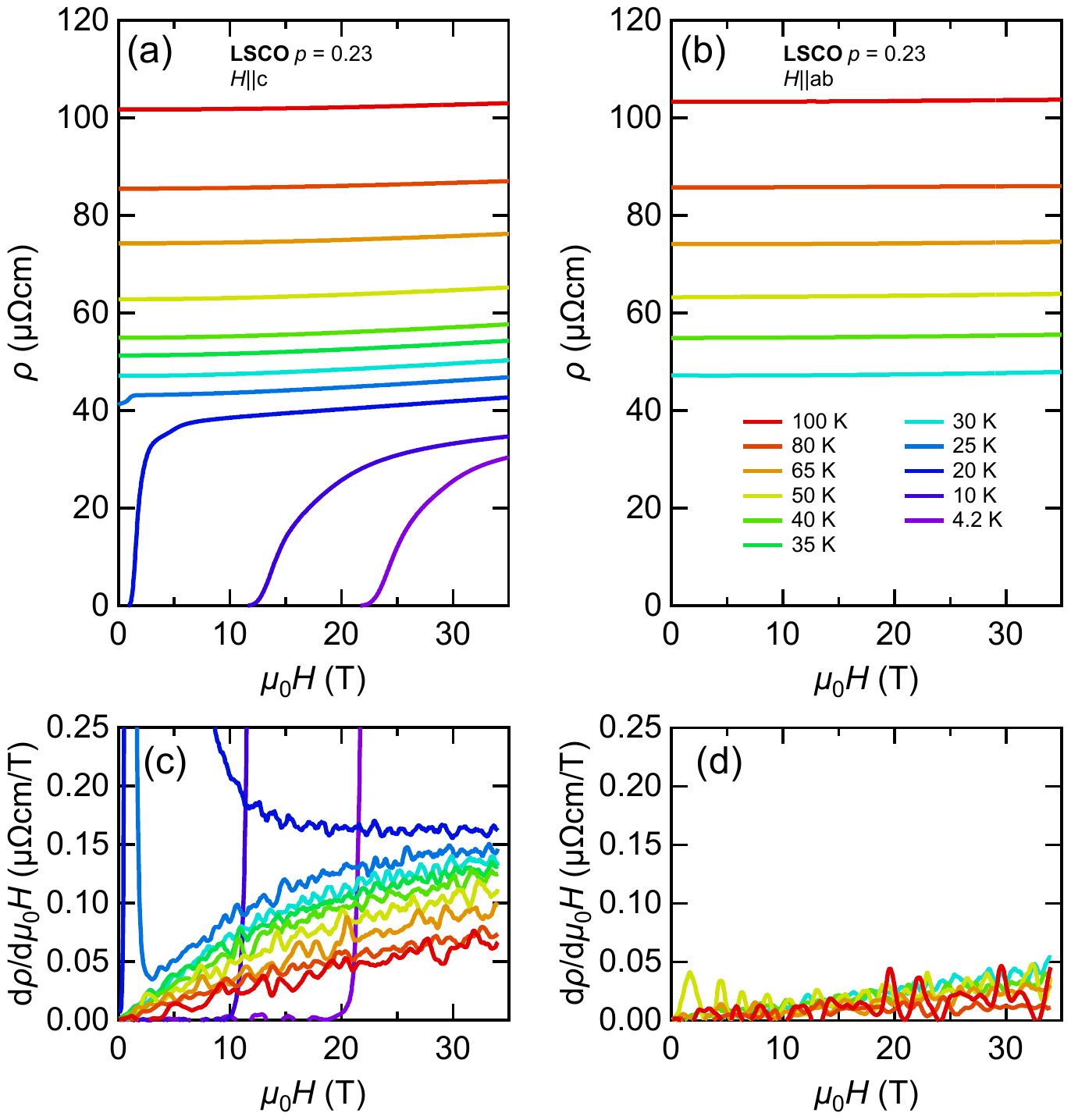}
         \caption{(a) Evolution of $\rho(H, T)$ for $p$ = 0.23 between 100 K and 4.2 K in magnetic fields up to 35 T oriented \textbf{H}$\|$$c$. (b) Evolution of $\rho(H, T)$ over the same range of temperatures and fields \textbf{H}$\|$$ab$. (c) Corresponding derivatives d$\rho$/d$\mu_0H$ for \textbf{H}$\|$$c$. (d) Corresponding d$\rho$/d$\mu_0H$ curves for \textbf{H}$\|$$ab$. The strong anisotropy in the two field orientations is in contrast to what was found for overdoped Bi2201 and Tl2201 \cite{Ayres_2020}.}
              \label{fig:LSCO_anisotropy}
\end{figure}

\end{document}
